\documentclass{article}
\usepackage{arxiv}

\usepackage[utf8]{inputenc} 
\usepackage[T1]{fontenc}    
\usepackage{hyperref}       
\usepackage{url}            
\usepackage{booktabs}       
\usepackage{amsfonts}       
\usepackage{nicefrac}       
\usepackage{microtype}      
\usepackage{lipsum}

\usepackage{times}
\usepackage{epsfig}
\usepackage{graphicx}
\usepackage{amsmath}
\usepackage{amssymb}
\usepackage{mathrsfs}
\usepackage{subfigure}
\usepackage{enumitem}
\usepackage[switch]{lineno}
\usepackage{indentfirst}
\usepackage{multirow}
\usepackage{color}
\usepackage{makecell}

\title{Interpolation Variable Rate Image Compression}

\author{Zhenhong Sun, Zhiyu Tan, Xiuyu Sun\thanks{Corresponding author.}, Fangyi Zhang, Yichen Qian, Dongyang Li, Hao Li\\
	\\Alibaba Group, China\\\\
	\texttt{\{zhenhong.szh, zhiyu.tzy, xiuyu.sxy, zhiyuan.zfy,}\\
	\texttt{yichen.qyc, yingtian.ldy, lihao.lh\}@alibaba-inc.com}}

\begin{document}
	\maketitle
\begin{abstract}
  Compression standards have been used to reduce the cost of image storage and transmission for decades. In recent years, learned image compression methods have been proposed and achieved compelling performance to the traditional standards. However, in these methods, a set of different networks are used for various compression rates, resulting in a high cost in model storage and training. 
  Although some variable-rate approaches have been proposed to reduce the cost by using a single network, most of them brought some performance degradation when applying fine rate control. To enable variable-rate control without sacrificing the performance, we propose an efficient Interpolation Variable-Rate (IVR) network, by introducing a handy Interpolation Channel Attention (InterpCA) module in the compression network. 
  With the use of two hyperparameters for rate control and linear interpolation, the InterpCA achieves a fine PSNR interval of 0.001 dB and a fine rate interval of 0.0001 Bits-Per-Pixel (BPP) with 9000 rates in the IVR network. Experimental results demonstrate that the IVR network is the first variable-rate learned method that outperforms VTM 9.0 (intra) in PSNR and Multiscale Structural Similarity (MS-SSIM).
\end{abstract}

\section{Introduction}

\begin{figure}[t]
	\centering
	\includegraphics[scale=0.41]{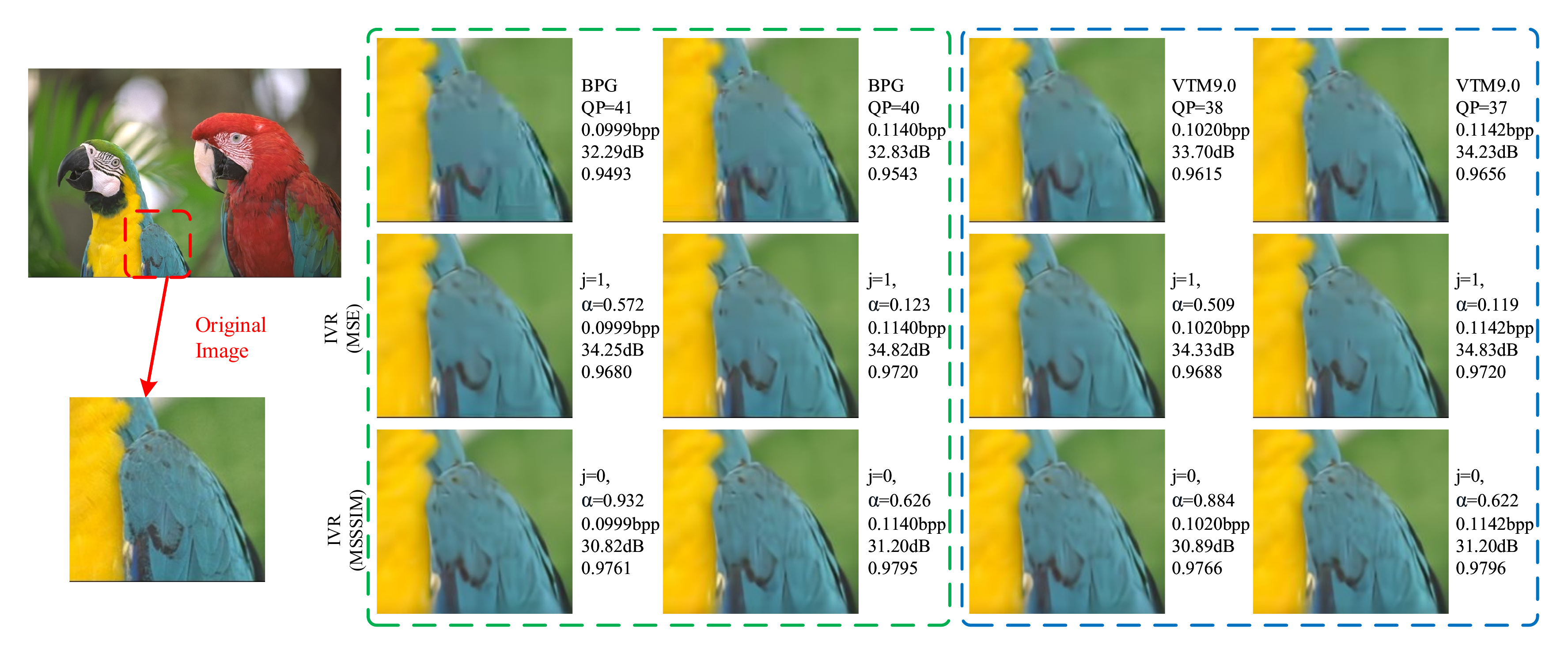}
	\caption{Visualization of sample images (Kodim23 from Kodak dataset) reconstructed by BPG, VTM 9.0, and the proposed IVR networks. The adjacent quantizer parameters (QPs) are used in BPG and VTM 9.0 to evaluate the fineness of their variable-rate control. By adjusting $j$ and $\alpha$, the IVR networks match the rates of BPG and VTM 9.0 with the fineness of 0.0001 BPP.}
	\label{fig:visual}
\end{figure}

Image compression aims to improve the efficiency of image storage and transmission by reducing data irrelevance and redundancy. In the past decades, many image compression standards have been proposed and widely used in the domain, such as JPEG~\cite{jpeg}, JPEG2000~\cite{jpeg2000}, AVC/H.264~\cite{AVC}, HEVC/H.265~\cite{HEVC}, etc. In recent years, with the development of deep learning and its advances shown in various computer vision applications~\cite{Druzhkov2016A,2019Deep}, more attention is paid to learned compression models via deep neural networks.

Inspired by the deep learning based transform coding~\cite{transform}, learned compression methods were proposed to use an entropy model to approximate the distribution of the compressible latents with CNNs~\cite{balle2017variational,theis2017lossy}, showing promising performance comparable to those traditional image codecs. On the basis of the entropy model, the performance of learned compression methods was further improved by introducing hierarchical structures, better entropy estimation models, and more appropriate network architectures~\cite{balle2018variational,minnen2019joint,lee2019,cheng2020learned,lee2019hybrid}. These methods quickly outperformed BPG (the intra image compression of HEVC) in terms of PSNR and MS-SSIM, among which the state-of-the-art method~\cite{lee2019hybrid} is still competitive to the recent VTM 9.0 (intra) -- the latest reference software of the VVC/H.266 standard~\cite{VVC}.

However, the aforementioned methods usually require training a set of separate networks for different compression rates, resulting in a high cost in model storage and training when multiple compression rates are desired in real applications. To reduce the cost, methods have been proposed to enable variable-rate control in a single network, by using a tuned loss function and additional hyperparameters (Lagrange multiplier and/or quantization step) to control the scale of the intermediate outputs or the latents~\cite{choi2019variable,yang2020variable,akbari2020learned}. While these methods significantly reduced the cost of model storage and training, they brought some performance degradation when applying fine rate control. There have been few works that achieve a good balance between the compression performance and the variable-rate control.

In this paper, we propose an efficient interpolation variable-rate network to enable variable-rate control without sacrificing the performance. The approach inserts InterpCA modules into the Encoder and Decoder of a baseline compression network without any modifications to the entropy model, reducing the risk of performance decline. The minimal plug-in design also makes the InterpCA module applicable to most types of entropy-based networks. The module has two crucial hyperparameters: (i) a rate hyperparameter to bind different rates with different settings in the InterpCA for coarse variable-rate control; (ii) an interpolation hyperparameter to interpolate the values of these settings for fine variable-rate control. With these two parameters, the IVR network can be trained with a range of sparse interpolation values to reduce the training cost, yet retain the variable-rate control with thousands of times finer rates during inference. This design is the key to the fine variable-rate control of the IVR method. 

Besides, optimizations are also made on the baseline autoregressive and hierarchical structure~\cite{minnen2019joint}. 
In the IVR model, most of the uniform-noise-addition operations are replaced by the quantization operations for latents to reduce the difference between training and inference. 
This design increases the accuracy of the entropy estimation and therefore improves the performance without increasing the computational complexity. 
In addition, a Unet post-network is also introduced to the pipeline in a modular manner to enhance the reconstruction. 
Benefiting from these optimizations, the IVR method is the first variable-rate learned method that outperforms VTM 9.0 (intra) in PSNR and MS-SSIM.

In particular, this paper has three major contributions:
\begin{itemize}[leftmargin=15pt,itemsep=2pt,topsep=2pt]
	\item[$\bullet$] By analyzing the histogram of latents, the principle of variable rates in learned image compression is explored, which inspires our design of the efficient variable-rate image compression.
	\item[$\bullet$] An interpolation method named as InterpCA is proposed to enable the variable-rate control without performance degradation. The minimal plug-in design of the InterpCA module makes it compatible with most entropy-model based methods.
	\item[$\bullet$] Comprehensive comparison and ablation studies are conducted to analyze the feasibility of the approach and how each component contributes to the performance improvement.
\end{itemize}

\begin{figure*}[htbp]
	
	\begin{minipage}{0.7\textwidth}
	\centering
	\includegraphics[scale=0.3]{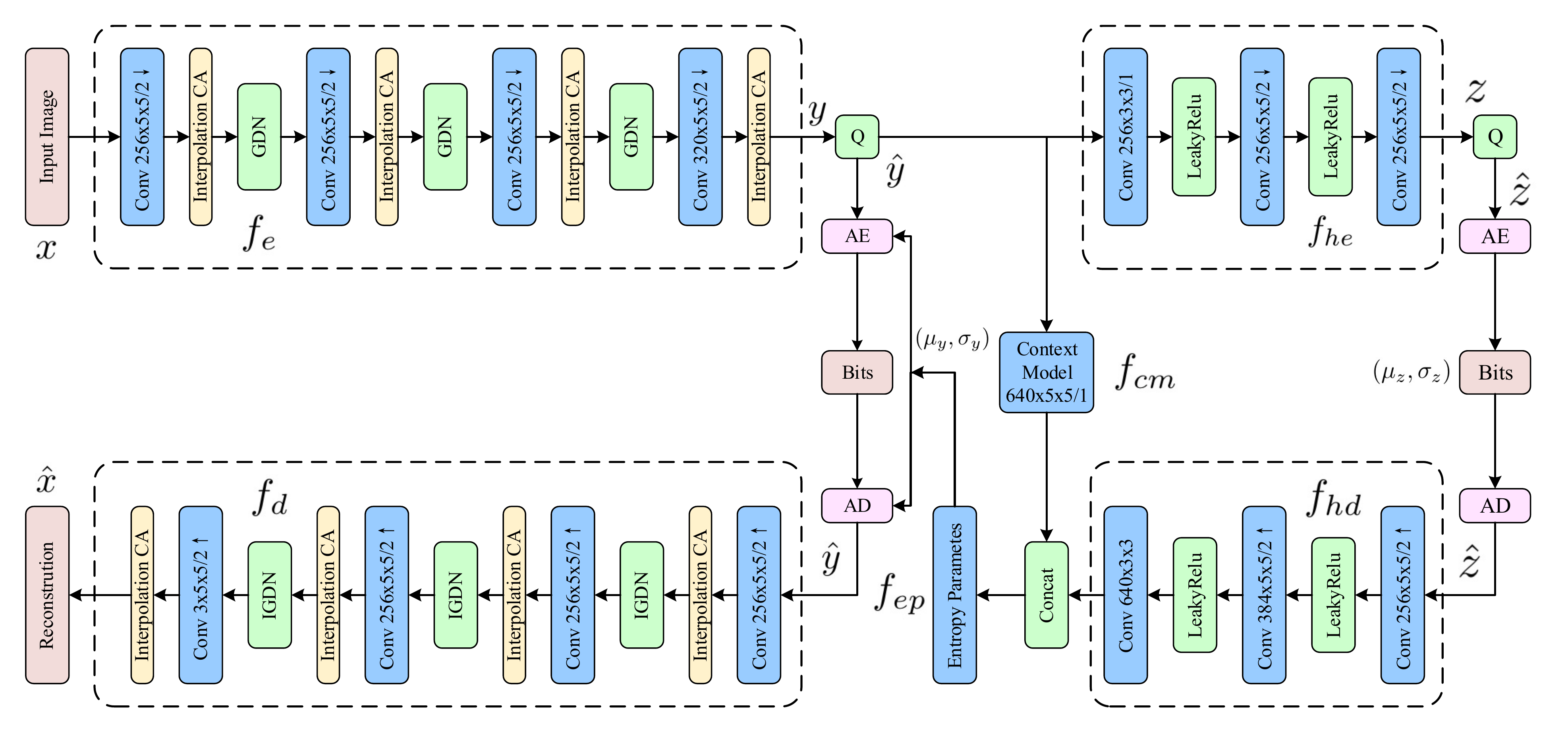}
	\end{minipage}
	\makeatletter
	\newcommand\tablecaption{\def\@captype{table}\caption}
	\makeatother
	\begin{minipage}{0.28\textwidth}
	\renewcommand\arraystretch{1.25}
	\centering
	\setlength{\abovecaptionskip}{0pt}%
	\setlength{\belowcaptionskip}{10pt}%
	\scalebox{0.65}{
	\begin{tabular}{c c}
		\hline
		Component & Symbol \\
		\hline
		Input Image & $x$\\
		Reconstruction & ${\hat{x}}$\\
		Encoder & $f_e(x;\theta_e)$\\
		Latent & $y$\\
		Latent(noised) & ${\tilde y}$\\
		Latent(quantized) & ${\hat y}$\\
		Decoder & $f_d({\hat y};\theta_d)$\\ 
		Hyper Encoder & $f_{he}({\hat y};\theta_{he})$\\
		Hyper Latent & $z$\\
		Hyper Latent(noised) & ${\tilde z}$\\
		Hyper Latent(quantized) & ${\hat z}$\\
		Hyper Decoder & $f_{hd}({\hat z};\theta_{hd})$\\  
		Context Model & $f_{cm}({\hat y_{mask}};\theta_{cm})$\\  
		Entropy Parameters & $f_{ep}(\cdot;\theta_{ep})$\\
		Unet Post-network & $f_{u}(\cdot;\theta_{u})$\\
		Post Reconstruction & ${\ddot{x}}$\\
		\hline

	\end{tabular}}
	\label{tab:symbol}
	\end{minipage}

	\caption{IVR Network architecture. AE/AD: Arithmetic Encoding/Decoding. Interpolation CA: Interpolation Channel Attention. Convolution parameters are denoted as the number of filters$\times$kernel height$\times$kernel width $/$ downsampling or upsampling stride, where $\uparrow$ indicates upsampling and $\downarrow$ downsampling. Ignoring the Interpolation CA, the remaining network structure is consistent with the structure of Minnen's autoregressive and hierarchical framework~\cite{minnen2019joint}.}
	\label{fig:network}
	\vspace{-0.4cm}
\end{figure*}

\section{Related Work}
\subsection{Learned Single Rate Image Compression}
Initially, recurrent neural networks were utilized in some works~\cite{toderici2015variable,toderici2017full,johnston2018improved} to recursively compress residual information using a binary representation to encode the latents of each iteration. Then the entropy-based method was proposed by Ball{\'e} \emph{et al.}~\cite{balle2017variational} and Theis \emph{et al.}~\cite{theis2017lossy}, which is the basis for the learned image compression methods. The entropy-based method consists of an Encoder to transform an image to a latent, an entropy model to reduce the entropy of the latent, and a Decoder to reconstruct the image from the latent, which resembles an autoencoder~\cite{autoencoder} with an entropy model. A hierarchical prior network was adopted in~\cite{balle2018variational} to enhance the entropy model by estimating the zero-mean Gaussian distribution of the latent representations. Mentzer \emph{et al.}~\cite{mentzer2018conditional} directly modeled the entropy of the latent representation by using a 3D-CNN Context Model. Since those spatially adjacent representations of latents have high correlations, Minnen \emph{et al.}~\cite{minnen2019joint} and Lee \emph{et al.}~\cite{lee2019} utilized context-adaptive entropy model with none zero-mean Gaussian distribution. Cheng \emph{et al.}~\cite{cheng2020learned} proposed an attention module to enhance the reconstruction and discretized Gaussian mixture likelihoods to improve the entropy model. Lee \emph{et al.}~\cite{lee2019hybrid} jointly optimized both the image compression and the quality enhancement with a Gaussian mixture model.

\subsection{Learned Variable Rate Image Compression}
Conditional Convolutions were proposed by Choi \emph{et al.}~\cite{choi2019variable} to realize a variable-rate image compression using the two-stage training. In the first training stage, the coarse variable-rate control was achieved by varying the Lagrange multiplier in the conditional model. While the fine variable-rate control was realized by tuning the quantization bin size of the latents in the second training stage. Yang \emph{et al.}~\cite{yang2020variable} proposed a similarly modulated autoencoder with different Lagrange multipliers to realize a coarse variable-rate control. In~\cite{akbari2020learned}, a B-bit quantizer with multiple values was introduced to realize the variable-rate image compression. Besides, to further improve the performance, the residual between the original image and the reconstructed image was encoded by BPG. 
Tong Chen and Zhan Ma proposed a set of quality scaling factors embedded after the Encoder network to achieve variable rates~\cite{chen2020variable}. 
These methods have achieved variable rates, but they don't consider the balance between the compression performance and the fine variable rates, which is vital to guarantee a good compression performance.

\section{Preliminary}\label{sec:s3}
\subsection{Autoregressive and Hierarchical Structure}
Figure~\ref{fig:network} shows the proposed IVR network architecture. If without the InterpCA (see Section~\ref{sec:IVR}), the rest parts in Figure~\ref{fig:network} compose the commonly used autoregressive and hierarchical framework~\cite{minnen2019joint} for image compression.
The Encoder transforms the input image $x$ into a latent $y$ by a transformation $f_e(x;\theta_e)$, which is then quantized into ${\hat y}$ for the next transformation. ${\hat y}$ can be losslessly compressed by arithmetic encoding (AE) and transmitted into a string of bits, using a probability distribution $p_{\hat y}({\hat{y})}$. A hyper-network (Hyper Encoder $f_{he}({\hat y};\theta_{he})$ and Hyper Decoder $f_{hd}({\hat z};\theta_{hd})$) with the Context Model and the Entropy Parameters network~\cite{minnen2019joint}, is utilized to learn the probability distribution of ${\hat y}$. The Context Model generates rough probability distribution parameters using a linear $5\times5$ masked convolution. Then, outputs of the Context Model and the Hyper Decoder are concatenated to generate accurate probability distribution parameters $({\mu_y}, {\sigma_y})$ by the Entropy Parameters network.

The learning goal of the entropy-based methods is to minimize the expected length of the bitstream as well as the expected distortion between the reconstructed and original images, 
leading to a Rate-Distortion (RD) optimization problem.
This can be formulated as:
\begin{equation}
\begin{split}
\mathbb{L} &= R+\lambda_j D \\
&=\mathbb{E}_{x\sim p_x}[-\log_2{p_{\hat y}({\hat{y})})}-\log_2{p_{\hat z}({\hat{z})})}]+\lambda_j \mathbb{E}_{x\sim p_x}{[d(x, \hat{x})]},
\end{split}
\label{eq:loss1}
\end{equation}
where $\lambda_j$ is the $j$th Lagrange multiplier $\lambda$ that determines the desired rate-distortion trade-off, $R$ is referred to as the expected length of the compressed bitstream, and $D$ is the distortion measured by either Mean Squared Error (MSE) or MS-SSIM. 
Similar to the traditional image codecs, the entropy model uses lossless entropy encoding to generate the final bitstream (such as arithmetic~\cite{arithmatic}), or lossless decoding algorithms to restore the latent from the bitstream (such as Huffman coding~\cite{huffman}). 
When multiple compression rates are desired in the single rate structure, these methods~\cite{minnen2019joint,cheng2020learned,lee2019,lee2019hybrid} normally require training a set of separate networks with different $\lambda$ for different compression rates, resulting in a very high cost in model storage and training. 

\begin{figure}[h]
	\centering
	\vspace{-0.4cm}
	\includegraphics[scale=0.4]{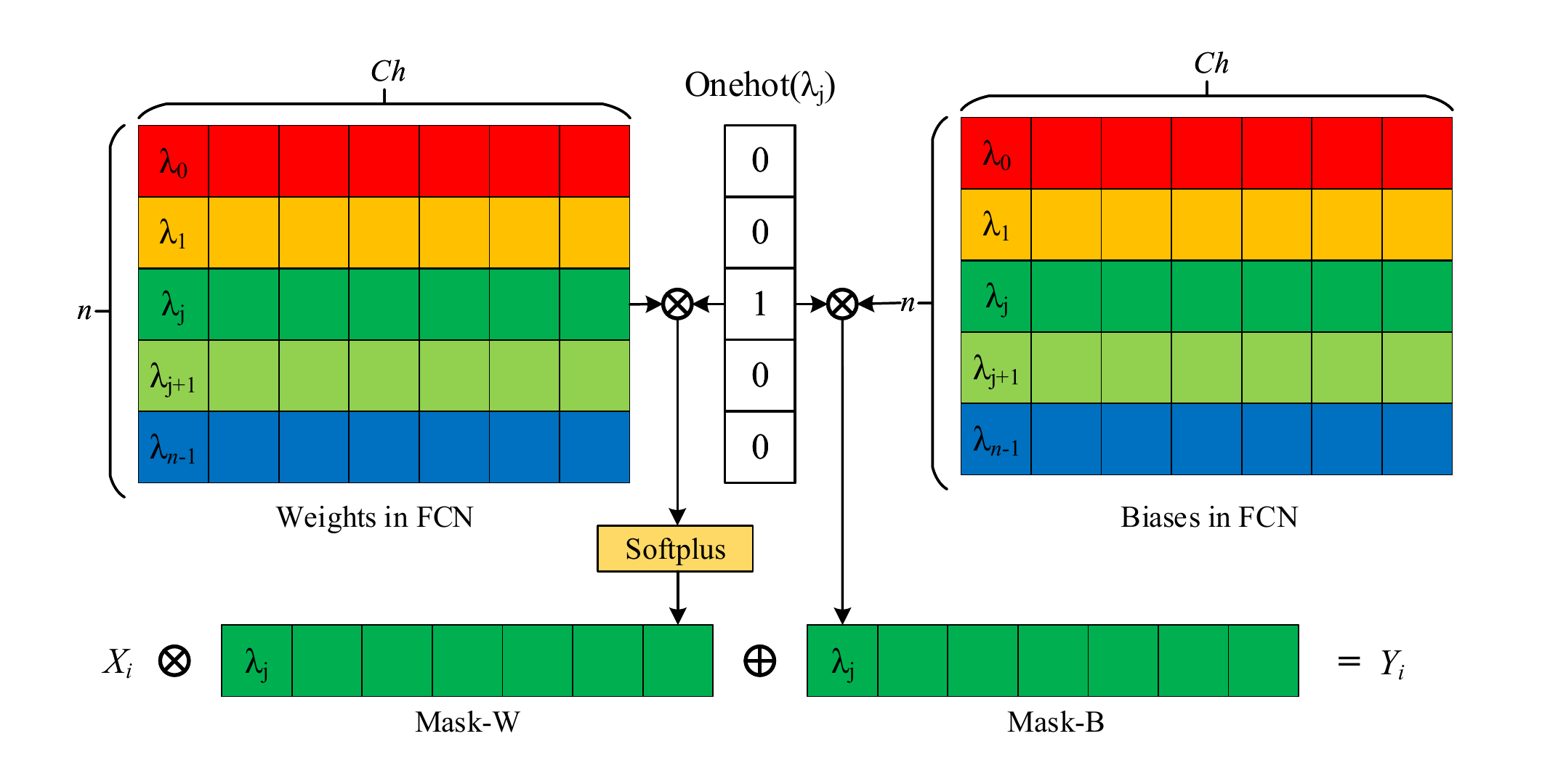}
	\vspace{-0.4cm}
	\caption{Conditional Convolution in Choi's paper~\cite{choi2019variable}. $n$ is the number of $\lambda$, $ch$ is the output channel of the convolution, $X_i$ is the output of the convolution, $Y_i$ is the output of the module. FCN: Fully-Connected Network.}
	\label{fig:CC}
	\vspace{-0.2cm}
\end{figure}

\begin{figure}[htbp]
	\centering
	\subfigure[$m(\lambda_j)$]{\includegraphics[scale=0.35]{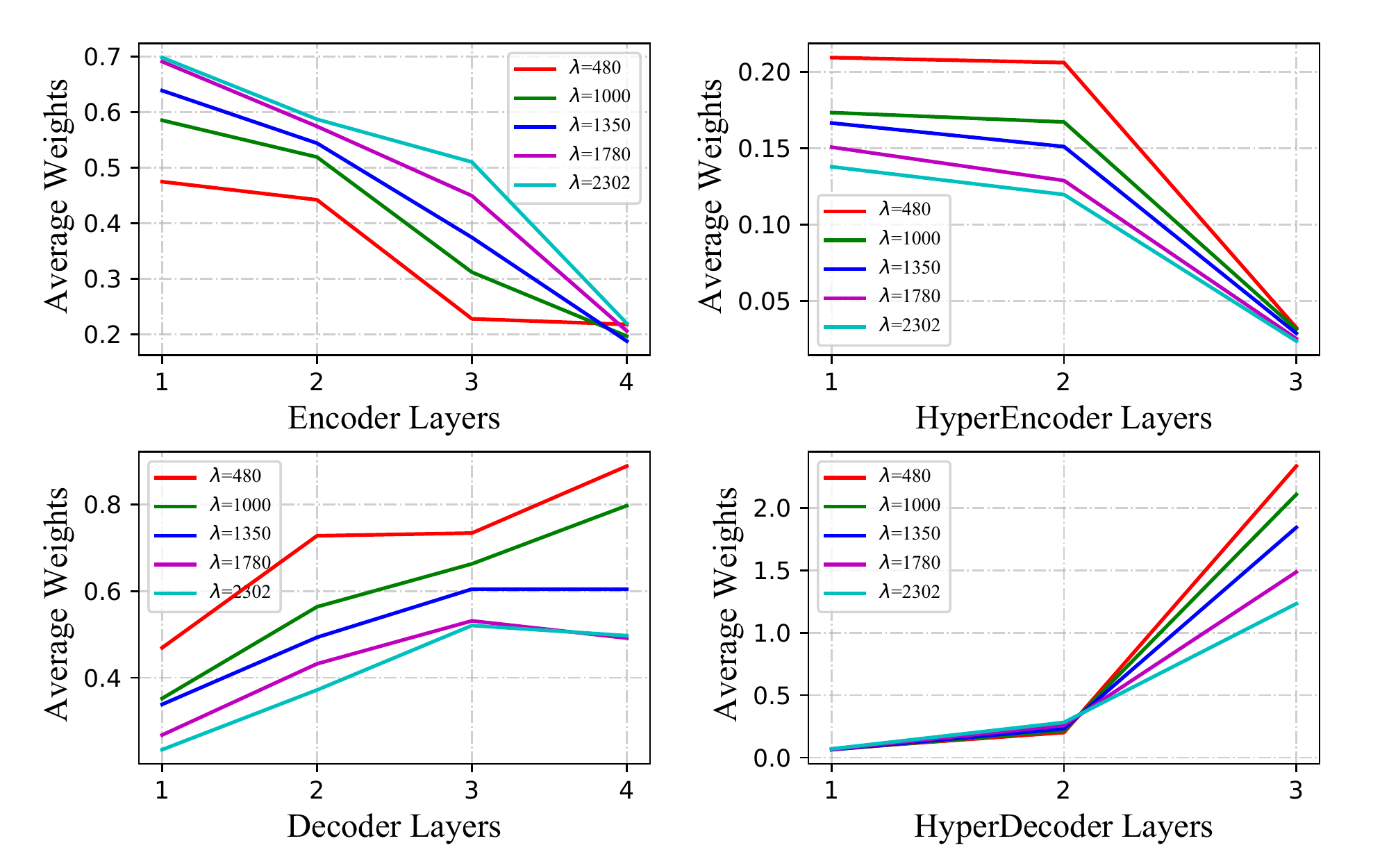} \label{sfig:weight}}
	\subfigure[$b(\lambda_j)$]{\includegraphics[scale=0.35]{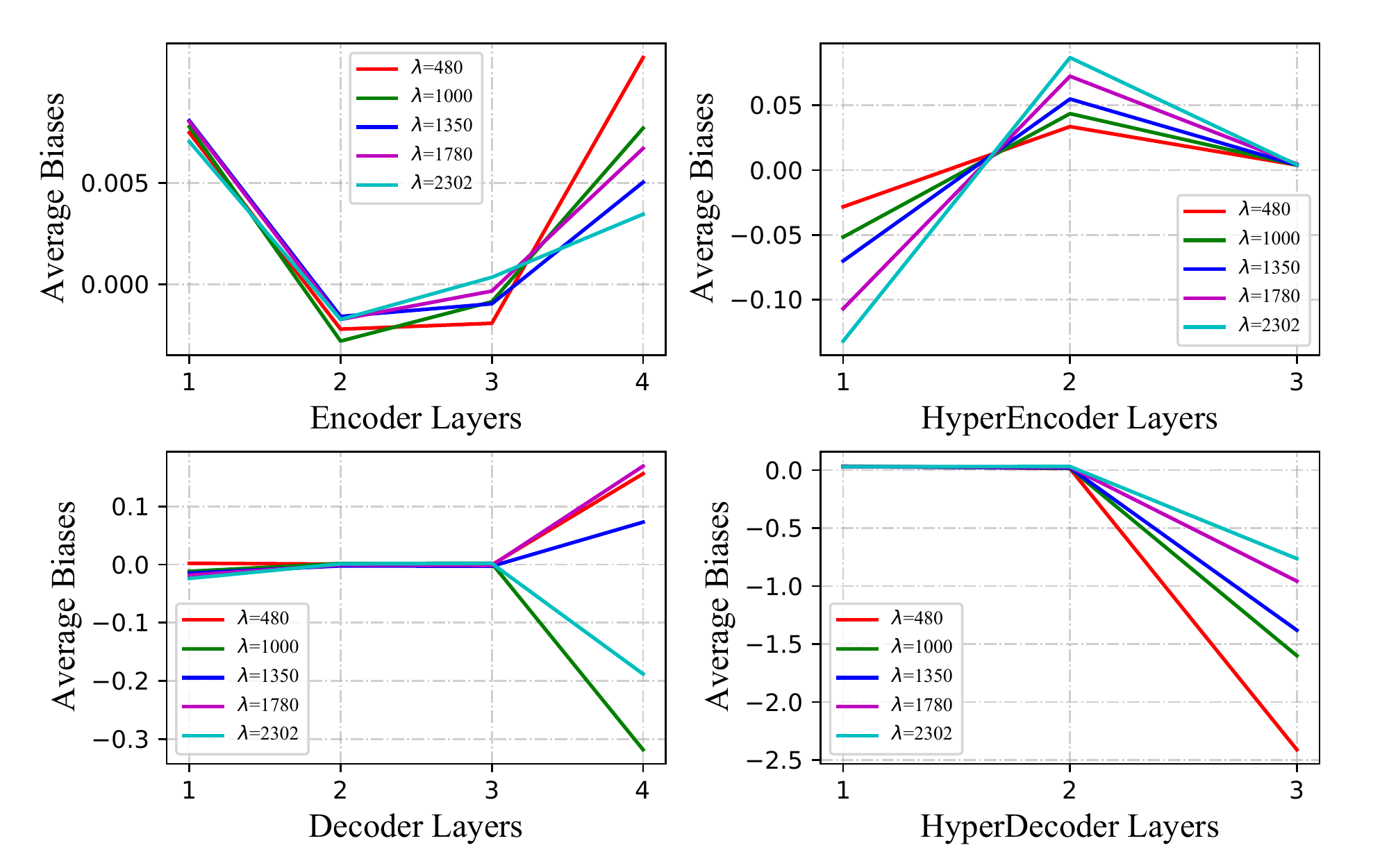} \label{sfig:bias}}
	\vspace{-0.4cm}
	\caption{The average values of $m(\lambda_j)$ and $b(\lambda_j)$ in different layers with different $\lambda$ in Conditional Convolution~\cite{choi2019variable}. The larger index corresponds to the rear layer of the network flow, \emph{e.g.}, the $4$ th layer in Decoder layers is the last layer of the network.}
	\label{fig:mask}
	\vspace{-0.2cm}
\end{figure}

\begin{figure}[htbp]
	\centering
	\vspace{-0cm}
	\subfigure[Histograms of $\hat y$]{\includegraphics[scale=0.36]{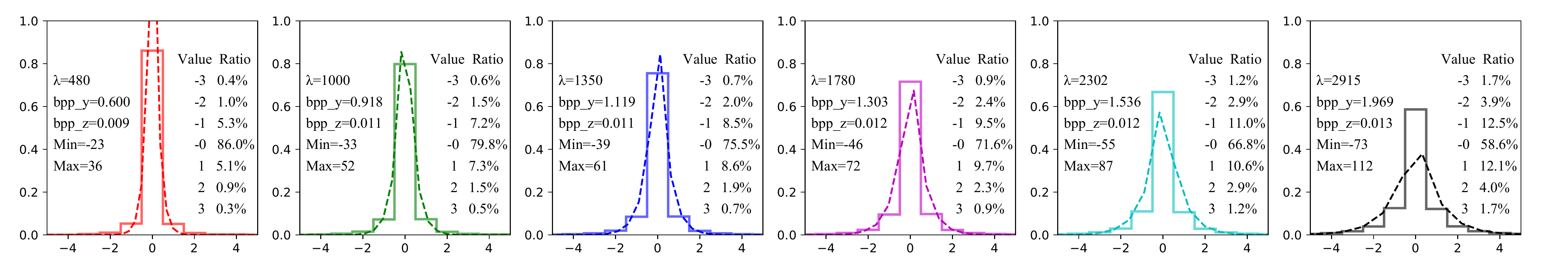} \label{sfig:y}}\vspace{-0.4cm}
	\subfigure[Histograms of $\hat z$]{\includegraphics[scale=0.36]{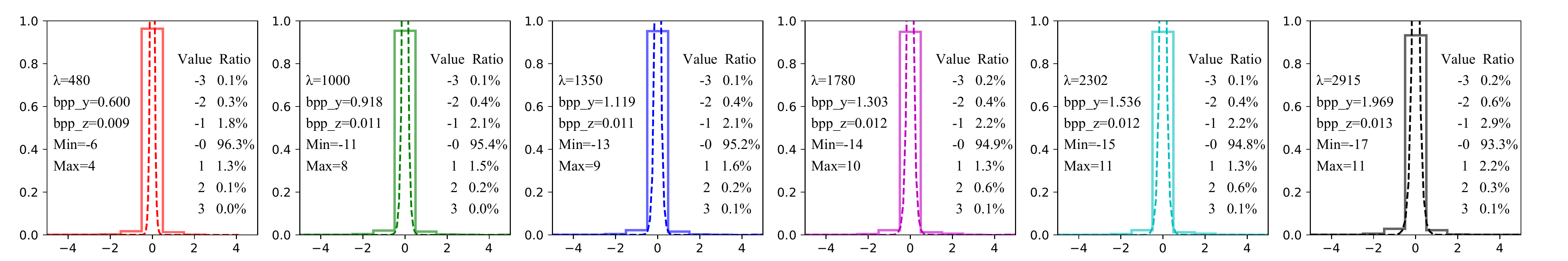} \label{sfig:z}}
	\vspace{-0.4cm}
	\caption{Histograms of $\hat y$ and $\hat z$ in Single Rate Network (SinRN) with different $\lambda$ on Kodim01. The solid line represents the histogram distribution, and the dotted line represents the Laplace fitting distribution.}
	\label{fig:Histogram}
	\vspace{-0.2cm}
\end{figure}

\subsection{Conditional Convolution}

The Conditional Convolutions~\cite{choi2019variable} realized the variable-rate control in an entropy-based single model. To figure out its principle, a detailed structure diagram is given in Figure~\ref{fig:CC}. Let $X_i$ be a 3-dimensional (3-D) feature map output of the $i$th convolution with channel $ch$,  and $Y_i$ be a 3-D feature map output of the Conditional Convolution with the same channel $ch$ as $X_i$. 
\begin{equation}
\begin{split}
Y_i&= m(\lambda_j) \otimes X_i + b(\lambda_j)\\
&= softplus\big(FCN(Onehot(\lambda_j)\big) \otimes X_i + FCN(Onehot(\lambda_j))\\
&\textup{with}\quad\lambda_j\in\{\lambda_0, \lambda_1, ..., \lambda_{n-2}, \lambda_{n-1}\}, 
\end{split}
\label{eq:lambja}
\end{equation}
where $j\in\{0,1,...,n-1\}$,  and $n$ is the number of pre-defined Lagrange multiplier values to control the variable rates. $m(\lambda_j)$ and $b(\lambda_j)$ are the channel-wised mask weights and biases for $X_i$.

In Choi's paper~\cite{choi2019variable}, all traditional Convolutions are replaced by the Conditional Convolutions to realize the coarse variable-rate control, while the fine variable-rate control was realized by tuning the quantization bin size of the latents. The compression performance on the Kodak dataset is wavy and incoherent (RD curves are shown in Figure~\ref{sfig:mse}), which leads to some performance degradation compared with Minnen's method~\cite{minnen2019joint}.

\subsection{Analysis of Variable Rates}

In the compression process, the entropy loss occurs in the Encoder to generate $\hat y$ and the process of Hyper compression is lossless to compress $\hat y$ with the AE~\cite{balle2018variational,minnen2019joint,lee2019,cheng2020learned,lee2019hybrid}. 
To verify the difference of variable rates in the Single Rate Networks (SinRNs), some experiments were conducted to display the distributions of latents with different $\lambda$ and Histograms of $\hat y$ and $\hat z$ are extracted and shown in Figure~\ref{fig:Histogram}. In the Histograms of $\hat y$, more values of input images are transformed to zero to reduce the whole entropy and the BPP of $\hat y$ decreases with the decrease of $\lambda$, so the SinRNs achieve different compression rates. Different from $\hat y$, the distributions of hyper-latents $\hat z$ remain nearly the same, indicating that the lossless Hyper compression might be independent of different compression rates.

Meanwhile, Conditional Convolutions~\cite{choi2019variable} were reproduced to evaluate the function of $m(\lambda_j)$ and $b(\lambda_j)$, and their average values of different layers are diplayed in Figure~\ref{fig:mask}.
Under fixed $\lambda$, the average weights decrease with the depth of the layers in Encoder Layers and HyperEncoder Layers, and increase with the depth of the layers in HyperDecoder Layers and Decoder Layers. Besides, the average weights are also affected by the value of $\lambda$. In particular, the average weights increase with $\lambda$ in the Encoder Layers and decrease in the other layers. 
Compared with the average weights of $m(\lambda_j)$, the average biases, most of which are valued around zero, do not exhibit apparent patterns. Considering this phenomenon, $b(\lambda_j)$ might be not necessary for Conditional Convolutions, whose removal could potentially help reduce the computational complexity. 

Similar to the situation of SinRNs, the Conditional Convolutions in the Encoder are controlled by $\lambda_j$ to produce $\hat y$ with different scaling coefficients, generating more zeros and realize a higher compression rate.
Theoretically, as long as the set of $\lambda$ is large enough and there are enough parameters in the Fully-Connected Network (FCN), it is possible for the model to obtain the fine variable-rate control.
However, as shown in Figure~\ref{fig:CC}, parameters in a row correspond to one $\lambda$ and they are independent of other rows of parameters during training and inference. If the set of $\lambda$ is large, each row of parameters may have fewer training iterations and be trapped in poor convergence, resulting in degraded compression performance. 
Additionally, the distributions of $\hat z$ are nearly the same as those in Figure~\ref{fig:Histogram}, hence the use of Conditional Convolutions might also bring performance degradation. 
Observing the red line ($\lambda=480$) and the green one ($\lambda=1000$) in Figure~\ref{sfig:weight}, if there exists another line between them, the model would be able to obtain an intermediate compression rate between $\lambda=480$ and $\lambda=1000$.
Based on these analyses, an interpolation variable-rate method is therefore proposed in this paper for a good balance between the compression performance and the variable-rate control.

\section{Proposed method}
\label{sec:IVR}
\subsection{Interpolation Variable Rate}

To solve the none correlations of parameters in Conditional Convolutions, a newly-elaborated method named InterpCA is proposed to bind different values of $\lambda$ with different rates by introducing a rate hyperparameter $j$ and an interpolation hyperparameter $\alpha$. The compression network takes $j$ and $\alpha$ as input hyperparameters that determine the value of $\lambda_{j,\alpha}$ according to the following equation:
\begin{equation}
\begin{split}
\lambda_{j,\alpha}&=\alpha \lambda_j+(1-\alpha) \lambda_{j+1}\\
&\textup{with} \quad {\lambda_j\ or\ \lambda_{j+1}} \in\{\lambda_0, \lambda_1, ..., \lambda_{n-2}, \lambda_{n-1}\}, 
\end{split}
\label{eq:ivr}
\end{equation}
where $j\in\{0,1,...,n-2\}$, $\alpha\in[0,1]$ and $n$ is the number of pre-defined Lagrange multiplier values.
Then, the loss function of Eq. (\ref{eq:loss1}) can be expressed as:
\begin{equation}
\mathbb{L} = R+\lambda_{j,\alpha} D.
\label{eq:loss2}
\end{equation}

\begin{figure}[h]
	\centering
	\vspace{-0.4cm}
	\includegraphics[scale=0.45]{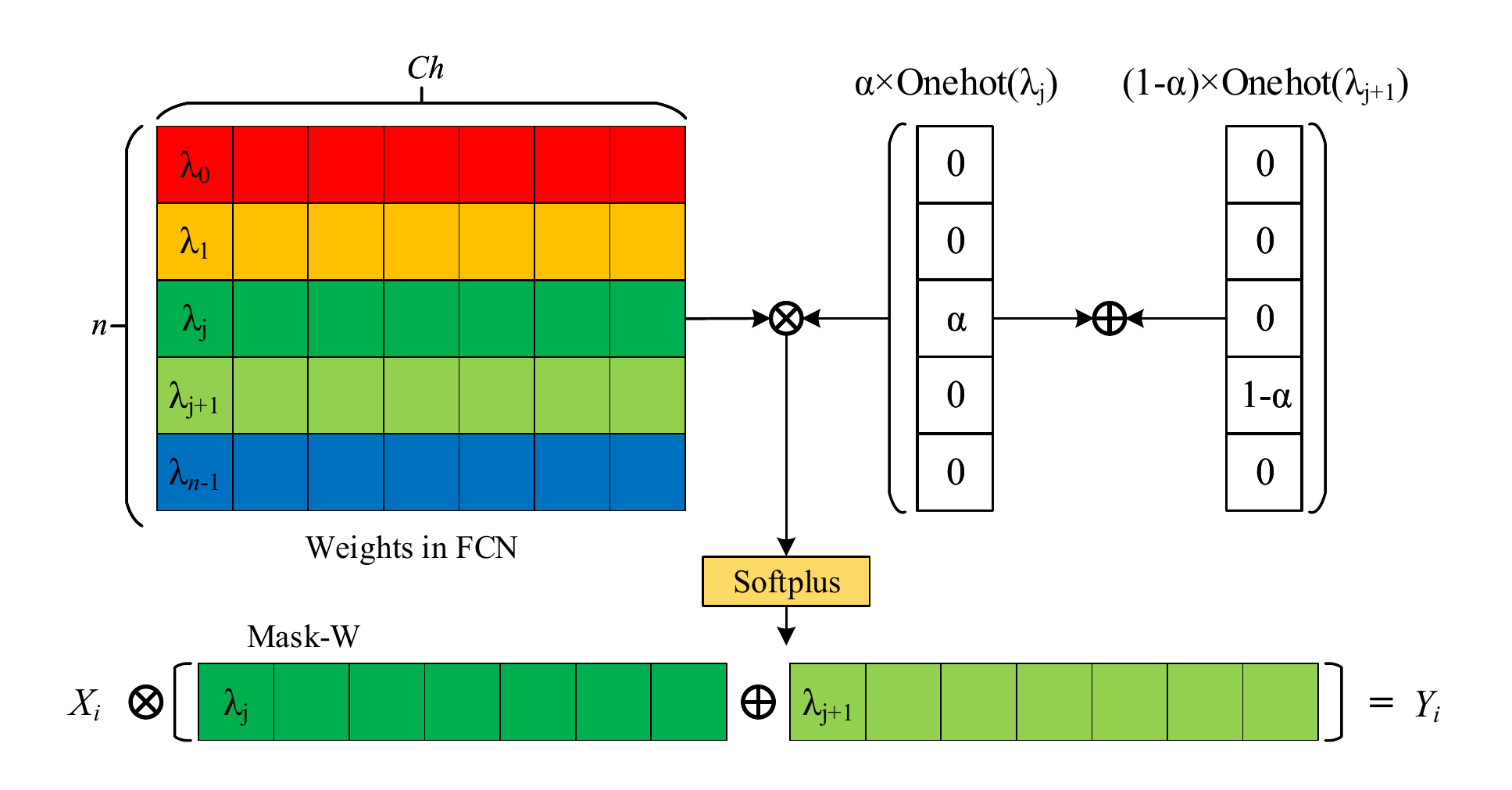}
	\caption{Interpolation channel attention (InterpCA). $n$ is the number of $\lambda$, $ch$ is the output channel of the convolution, $X_i$ is the output of convolution, $Y_i$ is the output of the module. In InerpCA, $j$ and $\alpha$ determine the final value of $\lambda$. Mask biases are not used in InterpCA.}
	\label{fig:ica}
	\vspace{-0.2cm}
\end{figure}

Figure~\ref{fig:ica} demonstrates the implementation of the InterpCAs, which are inserted behind each convolution layer in the Encoder $f_e(x;\theta_e)$ and the Decoder $f_d({\hat y};\theta_d)$. The output $X_i$ of each convolution layer can be transformed to $Y_i$  in the following way:
\begin{equation}
\begin{split}
Y_i &=softplus\big(FCN(I_n(j,\alpha))\big) \otimes X_i\\
&\textup{with}\quad I_n(j,\alpha)=\alpha I_n[j]+(1-\alpha) I_n[j+1], 
\end{split}
\label{eq:yi}
\end{equation}
where $I_n$ is a two-dimensional identity matrix showed, $\otimes$ denotes element-wised multiplication, $I_n[j/j+1]$ is the  $j/j+1$ th row vector of $I_n$ (\emph{i.e.}, $Onehot(\lambda_{j/j+1)}$), and the output shape of $softplus$ is [Batchsize, 1, 1, $ch$]. 

It is worth noting that the value range of $\alpha$ is different in training and inference. In each iteration of training, $j$ and $\alpha$ are randomly sampled within the value range (\emph{i.e.}, $\alpha\in\{0,0.5,1\}$) and then fed into the network. When $\alpha$ is equal to 0.5, $\lambda_{j,\alpha}$ is interpolated by $\lambda_j$ and $\lambda_{j+1}$ as $\lambda_{j,0.5}=(\lambda_j+\lambda_{j+1})/2$. In inference, $\alpha\in\{0,1/M,...,M-1/M,1\}$ is used to make full use of the correlations between the parameters in fully connected neural layers to realize nearly continuous rate control. Compared with existing methods, the InterpCA method has obvious advantages. One of the assets of the technique is its simplicity and efficiency, which helps most of the entropy-based SinRNs upgrade to the variable-rate network. Another advantage is that the rough range of interpolation hyperparameter enhances the robustness of the network and speeds up the convergence in training, and the superfine rate control could be realized by selecting an appropriate $M$ in inference. There are two reasons why InterpCA does not bring performance degradation. One is that inserting InterpCA into the Encoder and Decoder brings no change to the entropy model, and the other is that the InterpCA makes full use of the parameters in FCNs.

\subsection{Improved Single Rate Network}
\begin{figure}[h]
	\centering
	\vspace{-0.4cm}
	\includegraphics[scale=0.3]{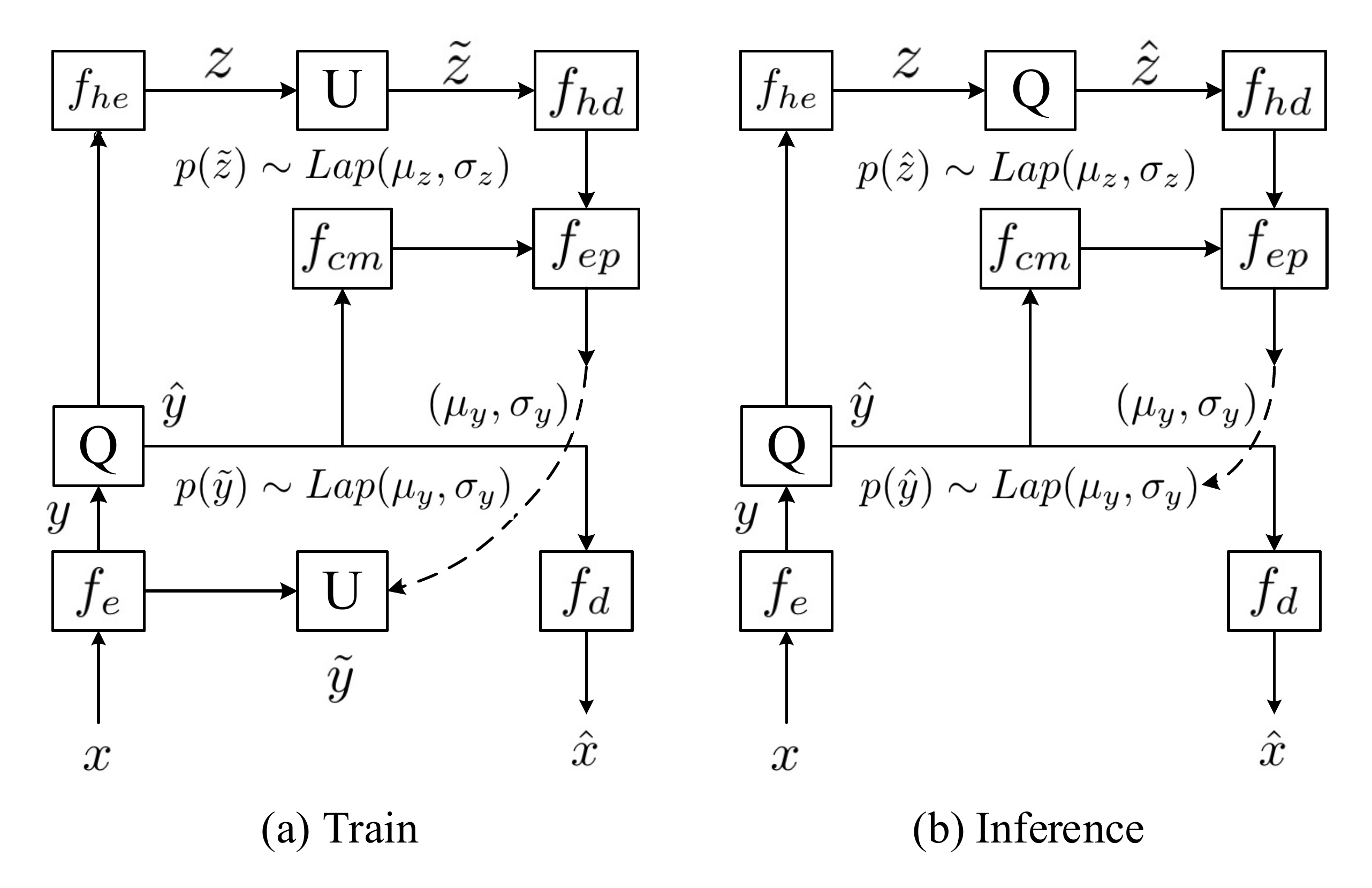}
	\vspace{-0.4cm}
	\caption{Operational Diagrams of SinRN. $Q$ denotes the operation of quantization. $U$ denotes the operation of adding uniform noise. The relationships between other symbols and components are summarized in Figure~\ref{fig:network}.}
	\label{fig:framwork}
	\vspace{-0.2cm}
\end{figure}

Figure~\ref{fig:framwork} shows the operation diagrams of the improved SinRN. 
Compared with~\cite{minnen2019joint}, most of the uniform-noise-addition operations are replaced by the quantization operations for latents in training that reduce the difference between the training and inference.
Similar to~\cite{Zhou_2019_CVPR_Workshops}, the probability $p_{\hat y}(\hat{y})$ of quantized latent ${\hat y}$ is modeled as Laplacian distributions: 
\begin{equation}
p_{\hat{y}}(\hat{y}|{\hat z}, \theta_{hd}, \theta_{cm}, \theta_{ep})=\prod_{i=1}\Big(\int_{{\hat{y}_i}-\frac{1}{2}}^{{\hat{y}_i}+\frac{1}{2}} Lap(y;\mu_y, e^{\sigma_y})\, dy\Big).
\label{eq:py1}
\end{equation}
To simulate the quantized process during training, we model each latent ${\tilde y}=y+U(-\frac{1}{2}, \frac{1}{2})$ as a Laplacian convolved with a unit uniform distribution. This ensures a good match between Encoder and Decoder distributions of both the quantized latents and continuous-valued latents subjected to additive uniform noise. The probability of ${\tilde y}$ is 
\begin{equation}
p_{\tilde{y}}(\tilde{y}|{\tilde z}, \theta_{hd}, \theta_{cm}, \theta_{ep})=\prod_{i=1}\Big(\int_{{\tilde{y}_i}-\frac{1}{2}}^{{\tilde{y}_i}+\frac{1}{2}} Lap(y;\mu_y, e^{\sigma_y})\, dy\Big).
\label{eq:py2}
\end{equation}

For the hyper latent $z$,  a set of channel-wise trainable parameters $(\mu_z, \sigma_z)$ are defined to learn the Laplacian distribution of the quantized latent ${\hat{z}}$ or noised latent ${\tilde{z}}$:
\begin{equation}
p_{\hat{z}}(\hat{z})=\prod_{i=1}\Big(\int_{{\hat{z}_i}-\frac{1}{2}}^{{\hat{z}_i}+\frac{1}{2}} Lap(z;\mu_z, e^{\sigma_z})\, dz\Big),
\label{eq:pz1}
\end{equation}
\begin{equation}
p_{\tilde{z}}(\tilde{z})=\prod_{i=1}\Big(\int_{{\tilde{z}_i}-\frac{1}{2}}^{{\tilde{z}_i}+\frac{1}{2}} Lap(z;\mu_z, e^{\sigma_z})\, dz\Big).
\label{eq:pz2}
\end{equation}
Particularly, transforming $\tilde{z}$ to the Hyper Decoder in the training period helps improve the entropy estimation of $p_{\tilde y}(\tilde{y})$ because additional uniform noise is beneficial for learning the probability distribution of ${\tilde y}$. The Decoder $f_d$ transforms $\hat{y}$ into the reconstruction $\hat{x}$. Because both the compressed latent and the compressed hyper latent are compressed by AE, the full loss function of the IVR network reads 
\begin{equation}
\begin{split}
\mathbb{L} &= R_y+R_z+\lambda_{j,\alpha} D \\
&=\mathbb{E}_{x\sim p_x}[-\log_2{p_{\tilde y}({\tilde{y})})} -\log_2{p_{\tilde z}({\tilde{z})})}] +\lambda_{j,\alpha} \mathbb{E}_{x\sim p_x}{[d(x, \hat{x})]}.
\end{split}
\label{eq:loss3}
\end{equation}\\
To further enhance the reconstruction like~\cite{lee2019hybrid}, a separable Unet post-network~\cite{Unet2015} shared by all rates is introduced to the model in this paper. Details of the Unet post-network are present in \textbf{\emph{Supplementary Materials}}.

\section{Experiments}
\subsection{Implementation Details}
\noindent\textbf{Details For Training} $\quad$
The networks were trained on a body of color PNG images licensed under creative commons, about 30K images downloaded from CVPR workshop CLIC training dataset~\cite{clic} and the world wide web. The networks were optimized using Adam with a batch size of 8 and a patch size of $256\times256$ randomly extracted from the training dataset. There were multistage learning rates ($\{1e-4, 5e-5, 1e-5, 5e-6, 1e-6\}$)  that changed with boundaries ($\{1600000, 2100000, 2300000, 2400000, 2500000\}$). 

The IVR network was optimized with two quality metrics, i.e., MSE and MS-SSIM. When optimized by MSE, $\lambda$ lain in $\{50, 160, 300, \\480, 710, 1000, 1350, 1780, 2302, 2915\}$ and $\alpha$ lain in $\{0, 0.5, 1.0\}$.  Different from MSE, $\lambda$  lain in $\{1, 2, 3, 5, 8, 10, 15, 20, 25, 30\}$, when optimized by $1-MS-SSIM$. Finally, the IVR network was enhanced by the Unet post-network to achieve better performance. Additional details about the network are shown in Figure~\ref{fig:network}.

\noindent\textbf{Details for Inference} $\quad$
We evaluated the compression performance on the commonly used 24 Kodak lossless images~\cite{kodak} with a size of 768 x 512 and 102 high-resolution CLIC validation images~\cite{clic}. For the IVR network, $M$ was set as 5 for drawing the whole RD curves and $M$ was set as 100/1000/1000 to evaluate the fine rate control at $j=8$. To evaluate the rate-distortion performance, the rate was measured by BPP, and the quality was measured by either PSNR or MS-SSIM, corresponding to the optimized distortion metric.

\noindent\textbf{Details For VVC and HEVC} $\quad$
We used the official test model VTM 9.0~\cite{VTM9} with intra profile and BPG software~\cite{BPG} to test the performance. For both of them, YUV444 format was used as the configuration to maximize the compression performance.

\begin{figure}[t]
	\centering
	\includegraphics[scale=0.26]{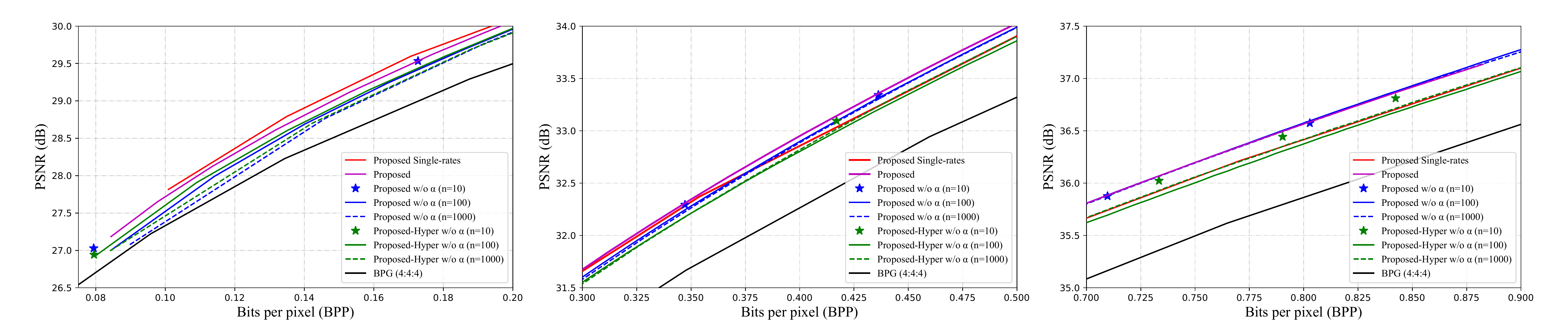}
	\caption{Ablation study of the IVR based on autoregressive and hierarchical structure~\cite{minnen2019joint} over the Kodak dataset. Three figures consist of the complete RD curves. IVR variants: "Proposed" is the baseline with InterpCA. "Proposed Single-rates" is the base SinRN. "Proposed w/o $\alpha$" represents the case without interpolation, which is realized by setting interpolation hyperparameter $\alpha$ to 1 in Eq.~\ref{eq:ivr}. "Proposed-hyper w/o $\alpha$" means the InterpCA is used in both Encoder/Decoder network and Hyper-Encoder/Decoder network with $\alpha=1$, which is similar to the case in~\cite{choi2019variable}.}
	\label{fig:ablation}
	\vspace{-0.3cm}
\end{figure}

\subsection{Ablation Study}
\begin{figure}
	\centering
	\includegraphics[scale=0.38]{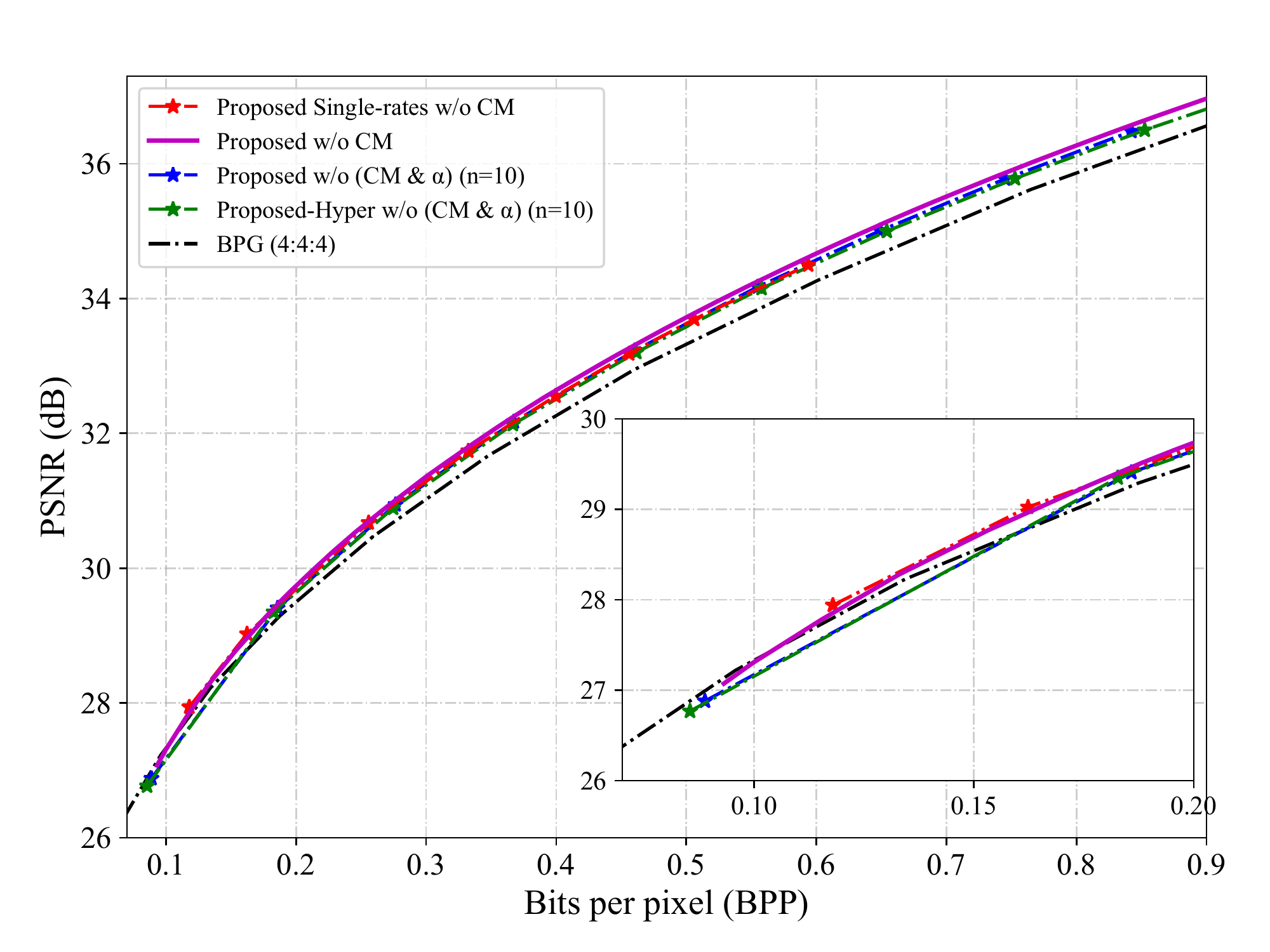}\vspace{-0.2cm}
	\caption{Ablation study of the IVR based on hierarchical structure~\cite{balle2018variational} over the Kodak dataset. "w/o CM" means the network is without the Context Model and the Entropy Parameters. Other variants are the same as Figure~\ref{fig:ablation}.}
	\label{fig:woCM}
	\vspace{-0.4cm}
\end{figure}

To verify the effectiveness of the interpolation variable-rate method, three kinds of IVR variants are implemented, \emph{i.e.}, \emph{the Proposed}, \emph{the Proposed without $\alpha$} and \emph{the Proposed-Hyper without $\alpha$}. \emph{the Proposed} is the baseline with InterpCA. \emph{the Proposed without $\alpha$} represents the case without interpolation, which is realized by setting interpolation hyperparameter $\alpha$ to 1 in Eq.~\ref{eq:ivr}. \emph{the Proposed-Hyper without $\alpha$} means the InterpCA is used in both Encoder/Decoder network and Hyper-Encoder/Decoder network with $\alpha=1$, which is similar to the case in~\cite{choi2019variable}. Meanwhile, these variants are deployed in two of the most common structures: autoregressive and hierarchical structure~\cite{minnen2019joint}, and hierarchical structure~\cite{balle2018variational}. These networks were all optimized by MSE, and results are summarized in Figure~\ref{fig:ablation}, Figure~\ref{fig:woCM} and Table~\ref{table:minnenbdbr}. Figure~\ref{fig:ablation} illustrates that compared with the SinRN for fixed $\lambda$, the proposed IVR network has no performance degradation on the whole and is better at some points.
The part A of Table~\ref{table:minnenbdbr} shows that the proposed model performs better than other IVR variants, saving 0.28\%, 2.28\%, 3.33\% bits (bits-saving is all measured by Bjøntegaard Delta Bit Rate (BDBR)~\cite{bdbr}) compared with SinRNs, \emph{the Proposed without $\alpha$ (n=1000)}, \emph{the Proposed-Hyper without $\alpha$ (n=1000)} on the Kodak dataset, respectively. 

The structure without $\alpha$ is equivalent to the Conditional Convolutions, the performance of \emph{the Proposed without $\alpha$ (n=1000)} deteriorates 1.93\% compared with the structure of 10 rates. This demonstrates that the compression performance decreases when using a large set of $\lambda$.
When inserting Conditional Convolutions into the Hyper Encoder and the Hyper Decoder, the performance of \emph{Proposed-Hyper without $\alpha$} drops 1.13\% (n=10), 1.47\% (n=100) and 1.05\% (n=1000), respectively. That means that sharing a common set of parameters in the Hyper-autoencoder network is sufficient for different compression rates.

\begin{table}[h]
	\begin{center}
		\caption{BD-Rate Gains against the SinRNs with and without the CM, corresponding to Part A and B respectively. Negative values in BDBR represent the bits saving.}
		\label{table:minnenbdbr}
		\vspace{-0.2cm}
		\renewcommand\arraystretch{1}
		\setlength{\abovecaptionskip}{0pt}%
		\setlength{\belowcaptionskip}{0pt}%
		\scalebox{0.8}{
			\begin{tabular}{|c|c|p{1.5cm}<{\centering}|p{2cm}<{\centering}|}
				\hline
				~&Methods & BDBR & BD-PSNR (dB)\\
				\hline
				\multirow{7}*{A}&Proposed &-0.2898\% & 0.0107\\
				~&Proposed w/o $\alpha$ (n=10) & 0.0592\% & -0.0040\\
				~&Proposed w/o $\alpha$ (n=100) & 1.3114\% & -0.0609\\
				~&Proposed w/o $\alpha$ (n=1000) & 1.9872\% & -0.0937\\
				~&Proposed-Hyper w/o $\alpha$ (n=10) & 1.1857\% & -0.0580\\
				~&Proposed-Hyper w/o $\alpha$ (n=100) & 2.7815\% & -0.1185\\
				~&Proposed-Hyper w/o $\alpha$ (n=1000) & 3.0409\% & -0.1340\\
				\hline
				\multirow{3}*{B}&Proposed w/o CM &-0.4740\% & 0.0193\\
				~&Proposed w/o (CM \& $\alpha$) (n=10) & 1.1733\% & -0.0482\\
				~&Proposed-Hyper w/o (CM \& $\alpha$) (n=10) & 1.6358\% & -0.0662\\
				\hline
		\end{tabular}}
	\end{center}
	\vspace{-0.4cm}
\end{table}

Different from the part serial compression of the autoregressive and hierarchical structure~\cite{minnen2019joint}, the hierarchical structure~\cite{balle2018variational} could be speeded up in parallel completely, which is more likely to be used in practice. The RD curves in Figure~\ref{fig:woCM} and BDBRs in the part B of Table~\ref{table:minnenbdbr} demonstrate the same conclusions as the ablation study based on the autoregressive and hierarchical structure. The part B of Table~\ref{table:minnenbdbr} shows that the proposed model without CM saves bits up to 1.65\%, 2.11\% compared with \emph{the Proposed without CM and $\alpha$ (n=10)}, and \emph{the Proposed-Hyper without CM and $\alpha$ (n=10)}. This indicates interpolation variable-rate method has more advantages in the hierarchical structure.

\begin{figure*}[!h]
	\centering
	\subfigure[PSNR on Kodak]{\includegraphics[scale=0.32]{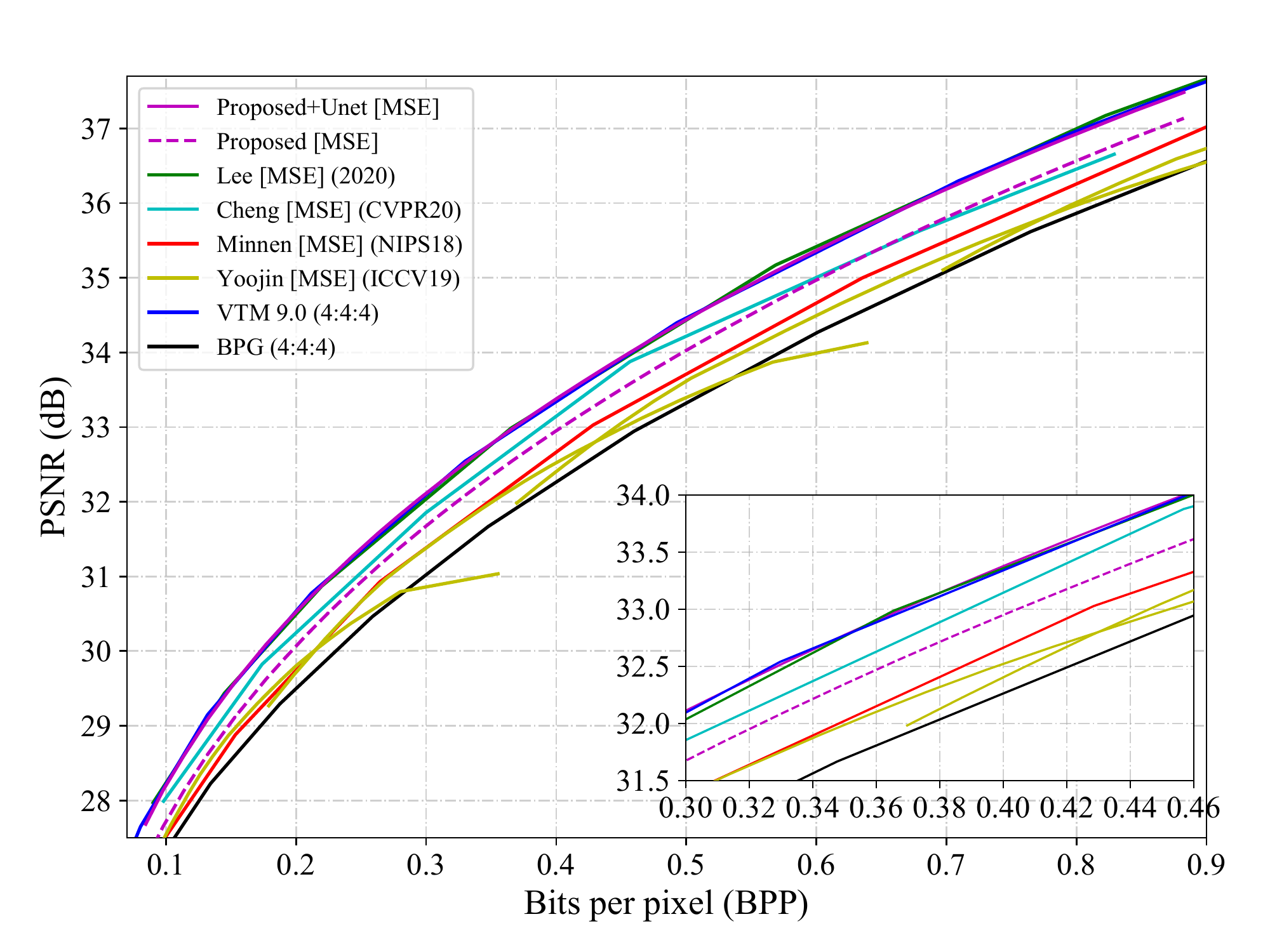} \label{sfig:mse}}
	\subfigure[MS-SSIM on Kodak]{\includegraphics[scale=0.32]{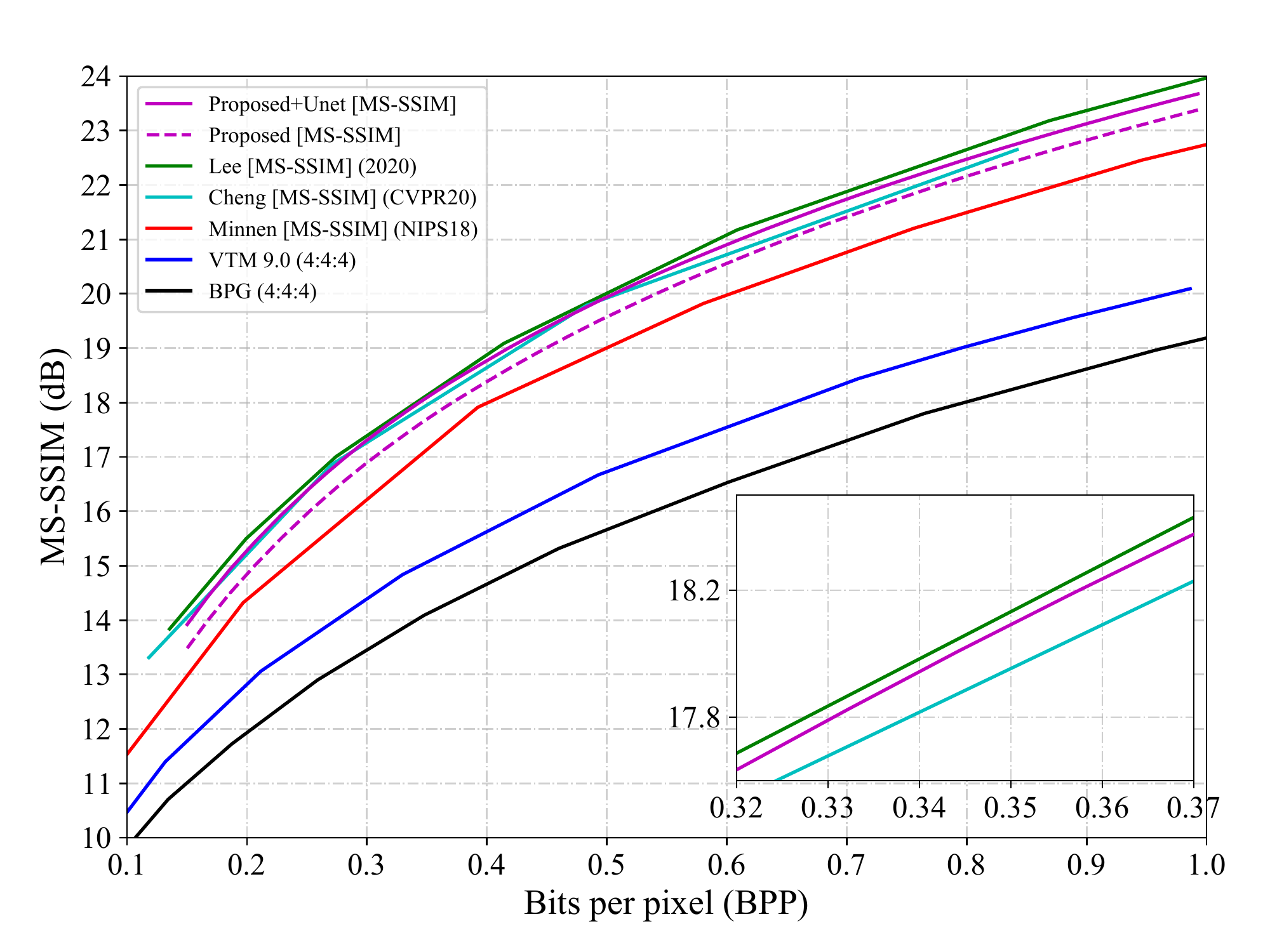}}\vspace{-0.4cm}
	\subfigure[PSNR on CLIC validation dateset]{\includegraphics[scale=0.32]{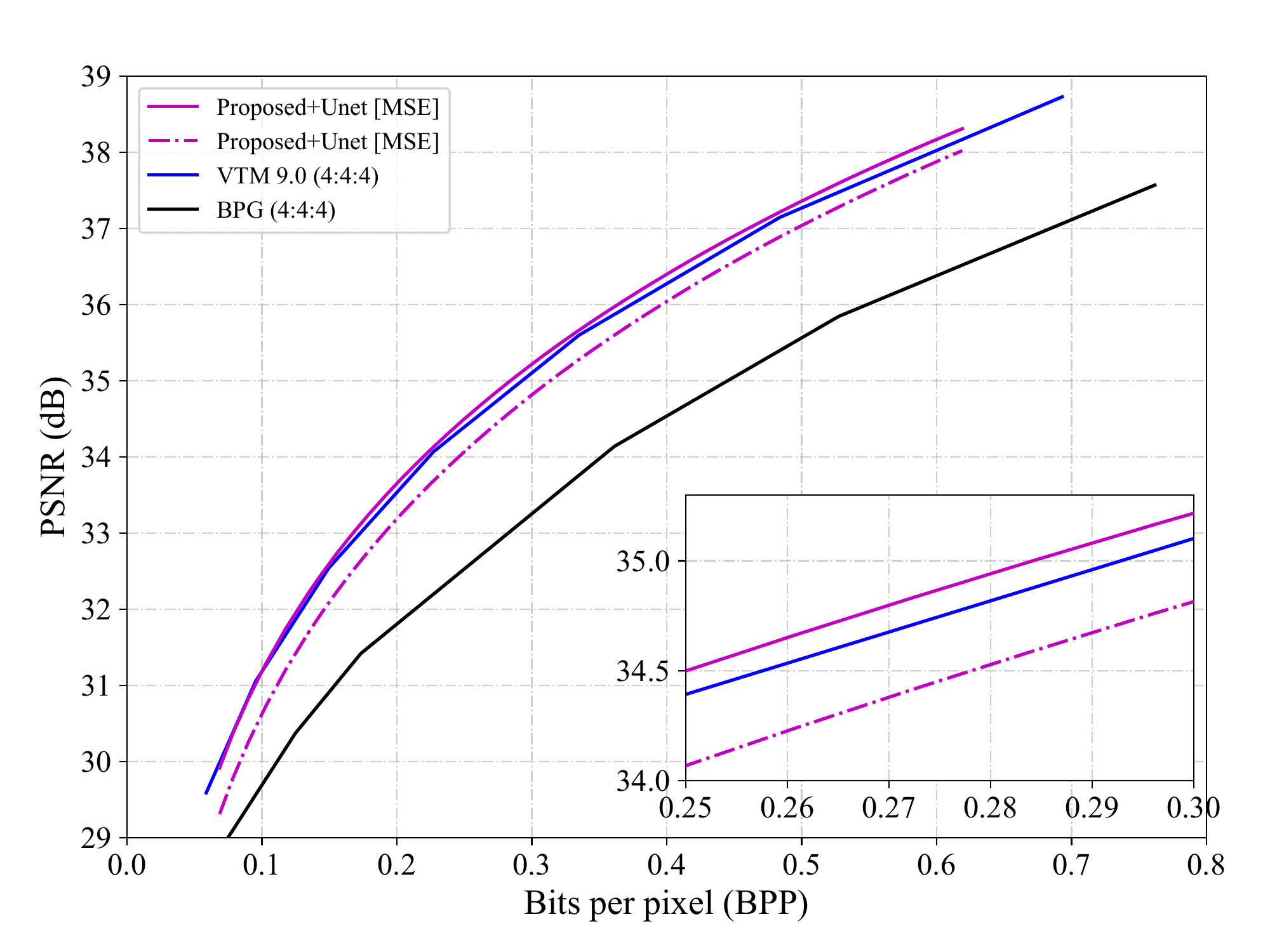}}
	\subfigure[MS-SSIM on CLIC validation dateset]{\includegraphics[scale=0.32]{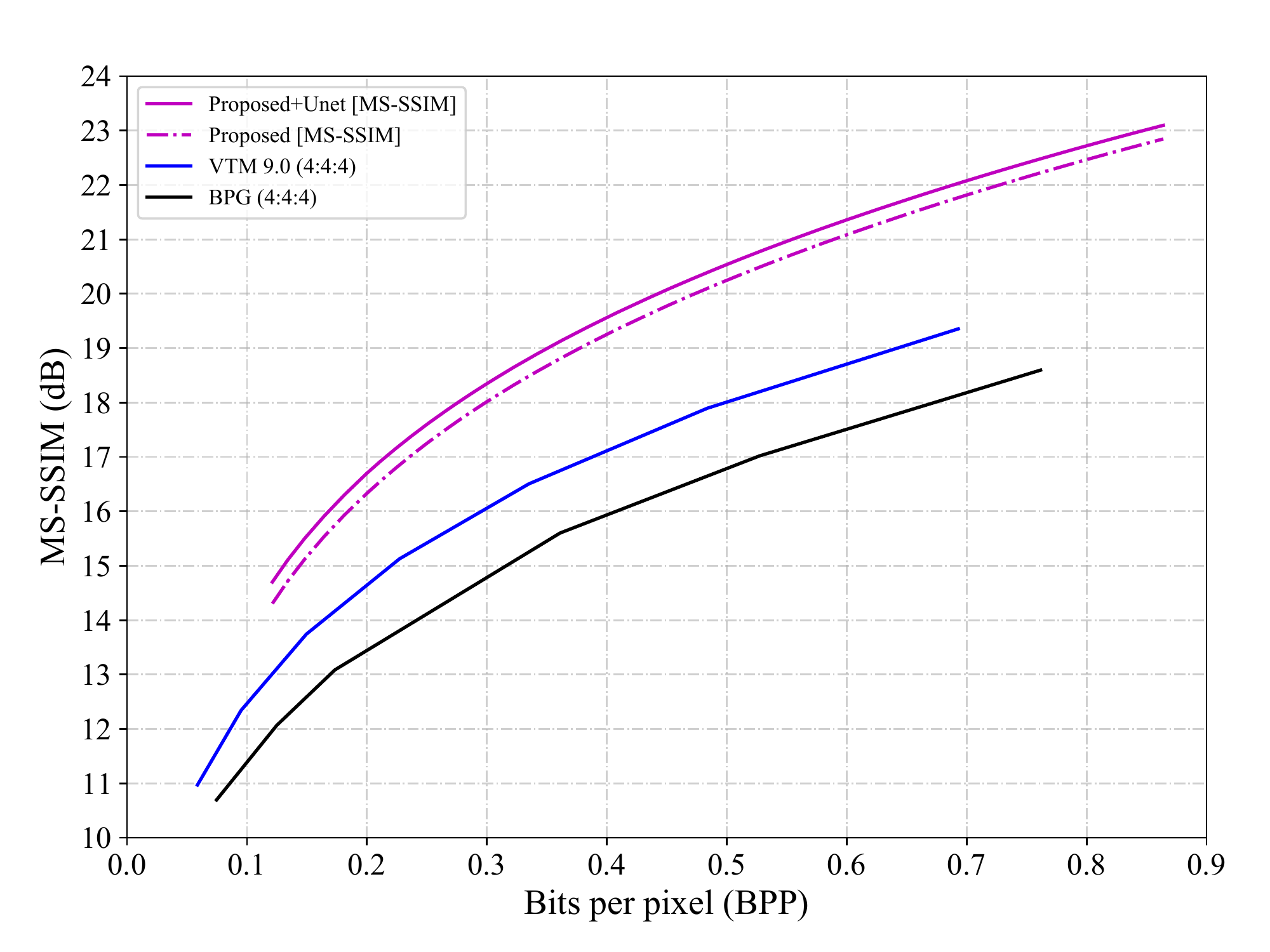}}\vspace{-0.4cm}
	\caption{RD curves aggregated over the Kodak and CLIC validation dataset. IVR-Unet (MSE) has a competitive rate-distortion performance on the Kodak image set as measured by PSNR (RGB) compared to all other methods. To our knowledge, this is the first learned variable-rate method that outperforms the VTM 9.0. MS-SSIM values converted to decibels $(-10log_{10}(1-MS-SSIM))$. The IVR-Unet (MS-SSIM) is slightly worse than~\cite{lee2019hybrid}.}
	\label{fig:RD}
	\vspace{-0.3cm}
\end{figure*}

\subsection{Rate Distortion Performance}
Figure~\ref{fig:RD} demonstrates the rate-distortion performance on the Kodak and CLIC validation dataset, where the IVR is compared with other previous methods, including well-known compression standards such as BPG and VVC, as well as recent entropy-based learned compression methods, such as Minnen's~\cite{minnen2019joint}, Cheng's~\cite{cheng2020learned}, Lee's~\cite{lee2019hybrid} and Choi's~\cite{choi2019variable}. RD curves of Minnen's~\cite{minnen2019joint} and Cheng's~\cite{cheng2020learned} are from the released LaTex code~\cite{cheng2020learned} in arXiv. RD curves of Lee's~\cite{lee2019hybrid} and Choi's~\cite{choi2019variable} are obtained by contacting the author. 
Regarding the metric of PSNR, the IVR network with Unet outperforms all other previous methods, achieving the state-of-the-art performance of its kind as shown in Figure~\ref{fig:RD} and Table~\ref{table:bdbr}. By setting the BPG as the anchor, the improvements of SinRN save $4.71\%$ bits than Minnen's~\cite{minnen2019joint} (Quantized operations contribute about 3.5\% and the Laplacian distribution contributes about 1\%) and Unet post-network boosts the performance about $8.40\%$ bits-saving. Although the improvements of SinRN promote the performance smaller than that of Unet post-network, it does not increase the computational complexity relative to Minnen's~\cite{minnen2019joint}.
Regarding the metric of MS-SSIM, it achieves a competitive performance among the learned image compression methods, which is only a little worse than Lee's~\cite{lee2019hybrid} and Cheng's~\cite{cheng2020learned}. It is also the first variable-rate image compression method that outperforms the ongoing compression method VTM 9.0 (intra) in both PSNR and MS-SSIM.
The results on the CLIC validation dataset show the IVR method also works for high-resolution images, apart from low-resolution Kodak images.

\begin{table}
	\begin{center}
		\caption{BD-Rate Gains of Proposed, Lee's~\cite{lee2019hybrid}, Cheng's~\cite{cheng2020learned}, VTM 9.0~\cite{VVC}, Minnen's~\cite{minnen2019joint}, against the BPG~\cite{BPG}. “/” represents that the method didn't evaluate the RD performance on the dataset in that column.}\vspace{-0.2cm}
		\label{table:bdbr}
		\renewcommand\arraystretch{1}
		\setlength{\abovecaptionskip}{0pt}%
		\setlength{\belowcaptionskip}{0pt}%
		\scalebox{0.8}{
			\begin{tabular}{|c|p{1.5cm}<{\centering}|p{1.5cm}<{\centering}|p{1.5cm}<{\centering}<{\centering}|p{1.5cm}<{\centering}|}
				\hline
				&\multicolumn{2}{c|}{Kodak} & \multicolumn{2}{c|}{CLIC}\\
				\cline{2-5}
				Methods & PSNR & MS-SSIM & PSNR & MS-SSIM\\
				\hline
				Proposed + Unet & -21.25\% & -56.36\% & -36.09\% & -58.31\%\\
				Proposed & -12.85\% & -52.71\% & -27.54\% & -54.26\% \\
				Lee's~\cite{lee2019hybrid} & -20.75\% & -57.98\% & /& /\\
				Cheng's~\cite{cheng2020learned} & -17.24\% & -57.38\% & /& /\\
				VTM 9.0~\cite{VVC} & -20.76\% & -20.13\% & -35.69\%& -28.43\%\\
				Minnen's~\cite{minnen2019joint} & -8.14\% & -47.948\% & /& /\\
				\hline
		\end{tabular}}
	\end{center}
	\vspace{-0.4cm}
\end{table}

We verify the rate fineness of the IVR-Unet network by changing the parameters $M$ with $j$ set as 8. The results in Table~\ref{tab:fvr} illustrate that the variable-rate RD points start to appear a disorder until $M = 10000$. In reality, traditional image codecs provide hundreds of variable-rate RD points to meet the basic requirement of applications. Compared with that, considering that the length of the values of $j$ is 9, the IVR network obtains 9000 effective variable-rate RD points with a very fine PSNR interval of 0.001 dB at $M=1000$. Moreover, as shown in Figure~\ref{fig:visual}, the IVR network realizes a fine rate interval of 0.0001 BPP compared with the traditional image codecs. Except for the variable-rate control, Figure~\ref{fig:visual} also reveals that the IVR image compression method provides better reconstruction quality with fewer artifacts in terms of PSNR, MS-SSIM, and perception.

\begin{table}[h]
	\renewcommand\arraystretch{1.25}
	\centering
	\setlength{\abovecaptionskip}{0pt}%
	\setlength{\belowcaptionskip}{0pt}%
	\caption{Fine variable-rate Test over the Kodak Dataset in IVR-Unet network (MSE).}\vspace{0.2cm}
	\label{tab:fvr}
	\scalebox{0.7}{
		\begin{tabular}{|c |c c |c c |c c|}
			\hline
			$j=8$&\multicolumn{2}{c|}{M=100} & \multicolumn{2}{c|}{M=1000} & \multicolumn{2}{c|}{M=10000}\\
			$\alpha$& BPP & PSNR(dB) & BPP & PSNR(dB) & BPP & PSNR(dB)\\
			\hline
			\setlength{\parskip}{10em}
			1 & 0.8023736 & 37.02175 & 0.8023736 & 37.02175 & 0.8023736 & 37.02175\\
			$1-1/M$ & 0.8032523 & 37.02725 & 0.8024466 & 37.02241 & 0.8023781 & 37.02176\\
			$1-2/M$ & 0.8040930 & 37.03298 & 0.8025299 & 37.02305 & 0.8023888 & 37.02185\\
			$1-3/M$ & 0.8049592 & 37.03910 & 0.8026113 & 37.02374 & 0.8023969 & 37.02196\\
			$1-4/M$ & 0.8058211 & 37.04532 & 0.8027042 & 37.02438 & 0.8024009 & 37.02206\\
			$1-5/M$ & 0.8066999 & 37.05148 & 0.8027963 & 37.02463 & 0.8024107 & 37.02214\\
			$1-6/M$ & 0.8076027 & 37.05822 & 0.8029035 & 37.02518 & 0.8024225 & 37.02213\\
			\hline
	\end{tabular}}
	\vspace{-0.4cm}
\end{table}

\section{Conclusion}

In this paper, we proposed an efficient IVR network for image compression by introducing an InterpCA module and using an improved SinRN. The original SinRN was optimized by replacing most of the uniform-noise-addition operations with quantization ones for more accurate entropy estimation, therefore bringing performance improvement without extra computational complexity.
A modular Unet post-network was also introduced to the pipeline to further enhance the reconstruction. Most importantly, benefitting from the InterpCA module which was used in both Encoder and Decoder of the SinRN, the IVR network can provide fine variable-rate control without performance degradation. 
In addition, the minimal plug-in design of the InterpCA makes it compatible with most entropy-based methods. As illustrated in Table~\ref{tab:fvr}, the proposed IVR network obtained 9000 effective variable-rate points with a fine PSNR interval of 0.001 dB and a fine BPP interval of 0.0001, when $M=1000$. The RD curves in Figure~\ref{fig:RD} validated that the IVR network outperformed VTM 9.0 (intra) in both PSNR and MS-SSIM. To the best of our knowledge, this is the first variable-rate learned image compression method achieving such competitive performance.


{\small
	\bibliographystyle{ref_fullname}
	\bibliography{reference}
}

\appendix
\newpage
\section{Details of Entropy Parameters}

\begin{figure}[!h]
	\centering
	\subfigure[Entropy Parameters module]{\includegraphics[scale=0.5]{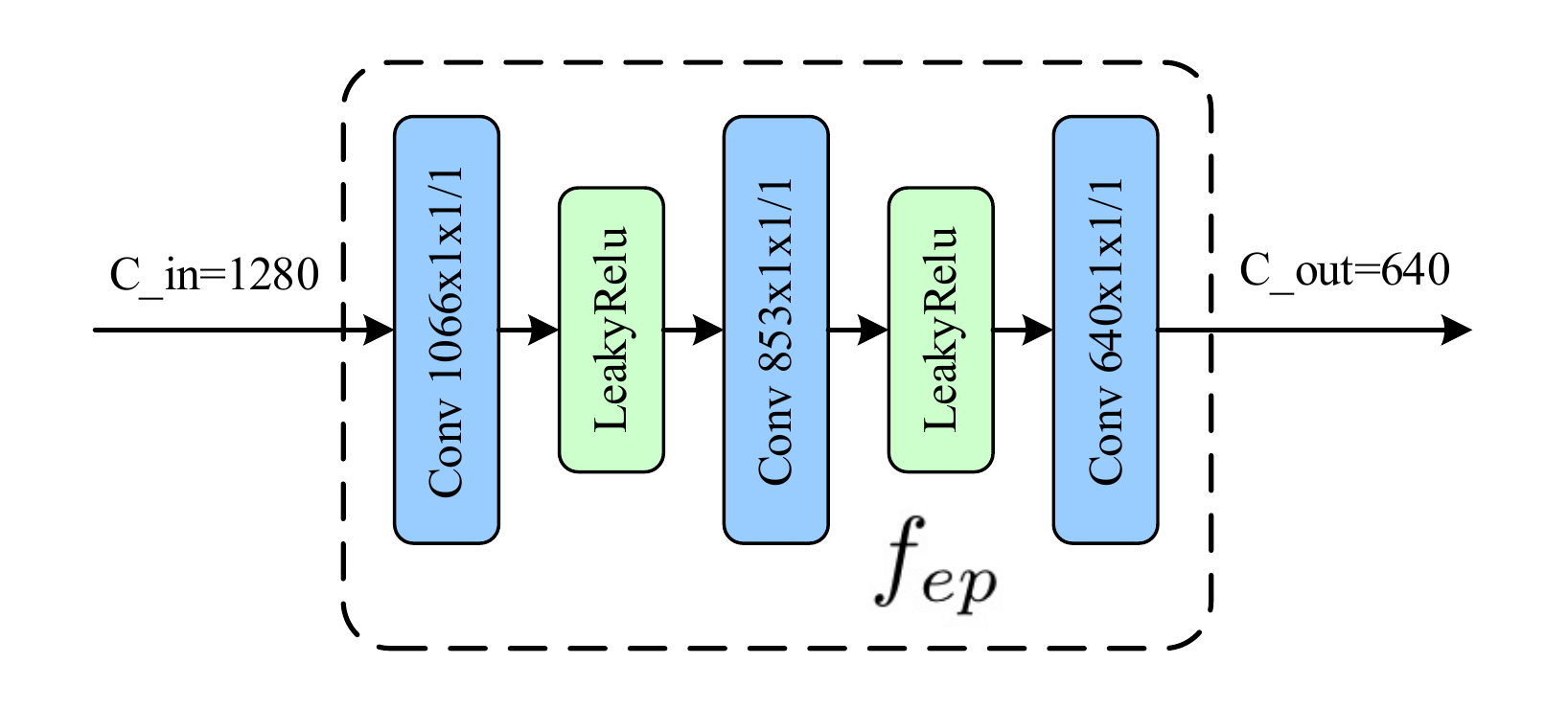}\label{sfig:epm}}
	\subfigure[Unet Post-network]{\includegraphics[scale=0.25]{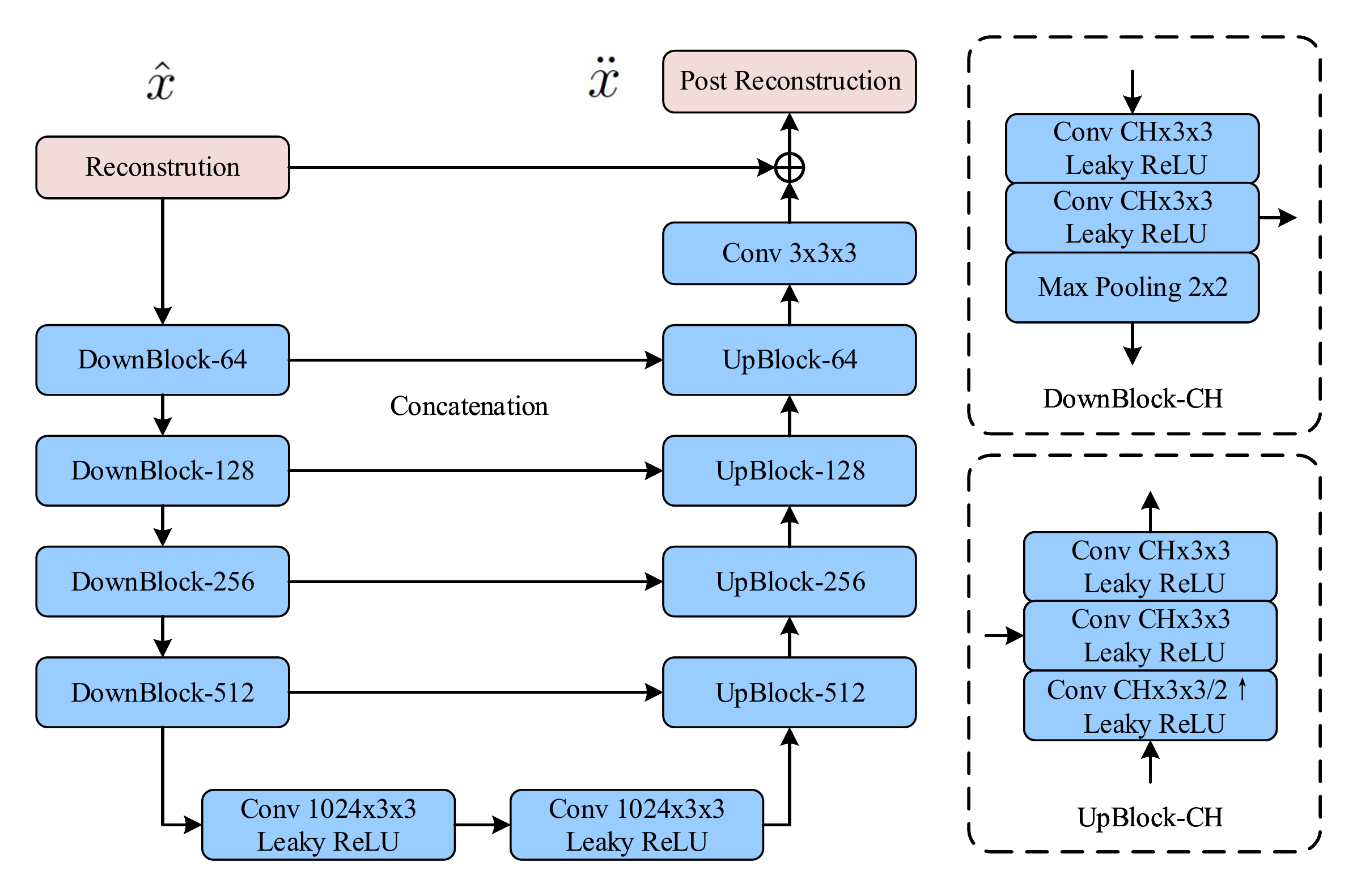}\label{sfig:upn}}
	
	\caption{Network architectures. Convolution parameters are denoted as the number of filters$\times$kernel height$\times$kernel width $/$ stride. Reconstruction $\hat{x}$ is down-up fourth in Unet structure to generate reconstruction $\ddot{x}$.}
	\label{fig:mmm}
	\vspace{-0.3cm}
\end{figure}

 The network architecture of the Entropy Parameters module is shown in Figure~\ref{sfig:epm}. The number of the input channels is equal to 1280, which is four times the channels of the latents ${\hat y}$. 
 Due to generating the probability distribution parameters $({\mu_y}, {\sigma_y})$, the final layer of the Entropy Parameters module has exactly twice as many channels as the latents ${\hat y}$.
 According to the channels of the input and the output, the number of $1\times1$ convolution channels decreases in a certain proportion ($ \frac{5}{6}, \frac{2}{3}, \frac{1}{2}$).

 \section{Details of Unet Post-network}

 To further enhance the reconstruction $\hat{x}$, a separable Unet post-network~\cite{Unet2015} is also introduced to the SinRN as shown in Figure~\ref{sfig:upn}. In the Unet post-network, all convolution layers are followed by a LeakyReLu, except for the last output layer. Combining this sub-network with the main network for end-to-end training can bring a little performance improvement~\cite{lee2019hybrid}. While the introduction of separated sub-networks has two advantages: (i) the separated sub-network can be replaced by other post-processing modules according to the actual needs, such as denoising, defogging, and so on, (ii)  the separated sub-network can also be removed to reduce the computational cost for the fast decoding.

 Compared with the baseline Minnen’s method~\cite{minnen2019joint}, Cheng’s method~\cite{cheng2020learned} saves 9.1\% bits with the complex Gaussian Mixture Model (GMM) entropy model and complex Encoder/Decoder, Lee’s method~\cite{lee2019hybrid} saves 12.61\% bits with a similar GMM entropy model and Grouped Residual Dense Network (GRDN), and IVR saves 13.11\% bits with improved entropy model and Unet.
 With the same input resolution of $1920\times1080\times3$, the main network of IVR has 28M parameters with 1.6 TFLOPs, the Unet has 32M parameters with 2.0 TFLOPs and saves 8.4\% bits. GRDN in Lee’s method~\cite{lee2019hybrid} has fewer parameters (5M, but they need to multiply N for N rates) with high FLOPs (3.0 TFLOPs) and saves 8.6\% bits (shown in their paper). So the Unet is used for a fair comparison with the SOTA Lee’s method, based on the same order of complexity of the Decoder.

 \section{Visualizations on Kodak}

\begin{figure*}[!hbp]
 	\centering
 	\includegraphics[scale=0.8]{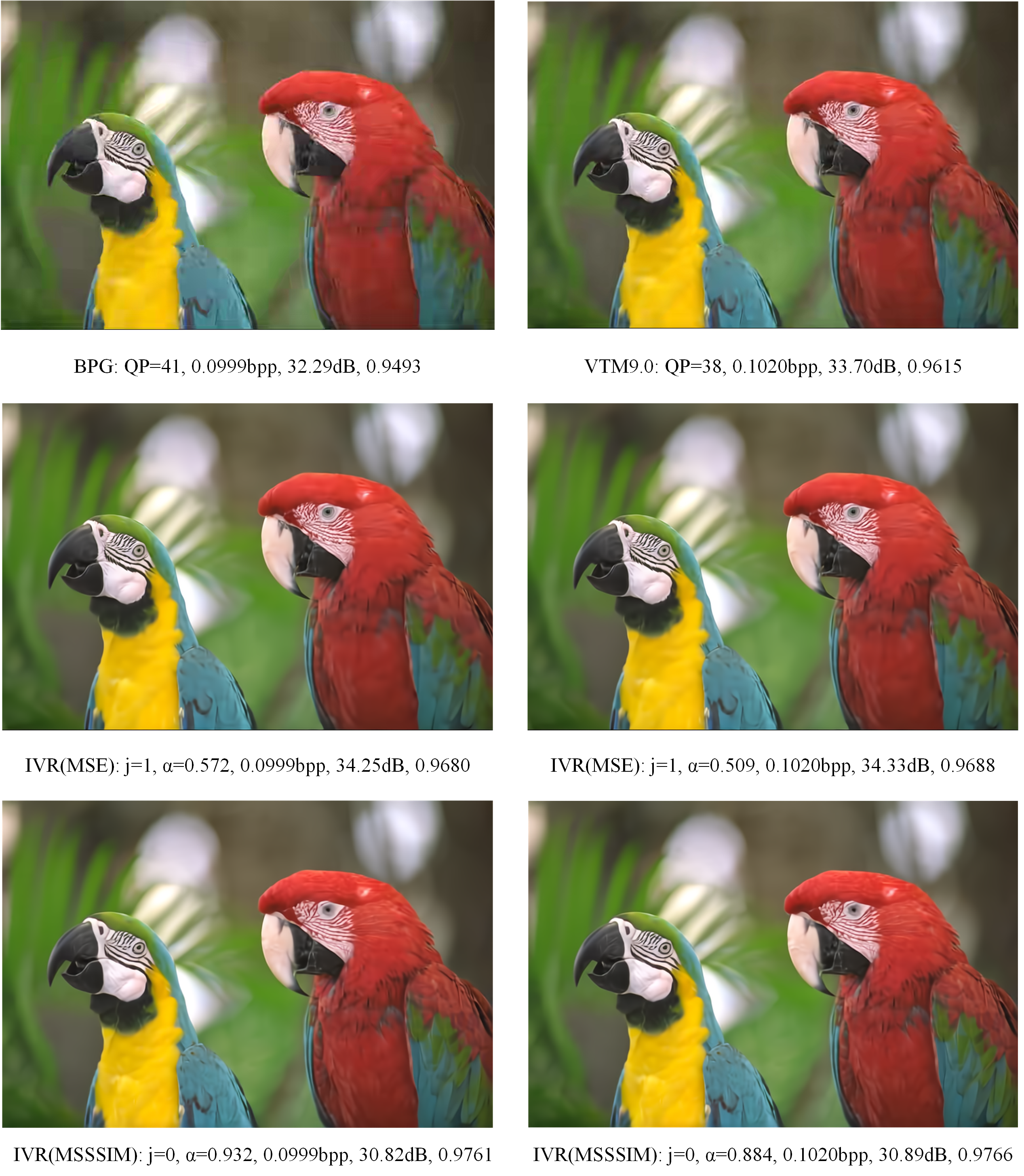}
 	\caption{Visualization of sample images (Kodim23 from Kodak dataset) reconstructed by our IVR networks, BPG and VTM 9.0. By adjusting $j$ and $\alpha$, the IVR networks (with Unet post-network) match the rates of BPG and VTM 9.0 with the fineness of 0.0001 BPP.}
 	\label{fig:visual1}
 \end{figure*}

\begin{figure*}[!hbp]
 	\centering
 	\includegraphics[scale=0.8]{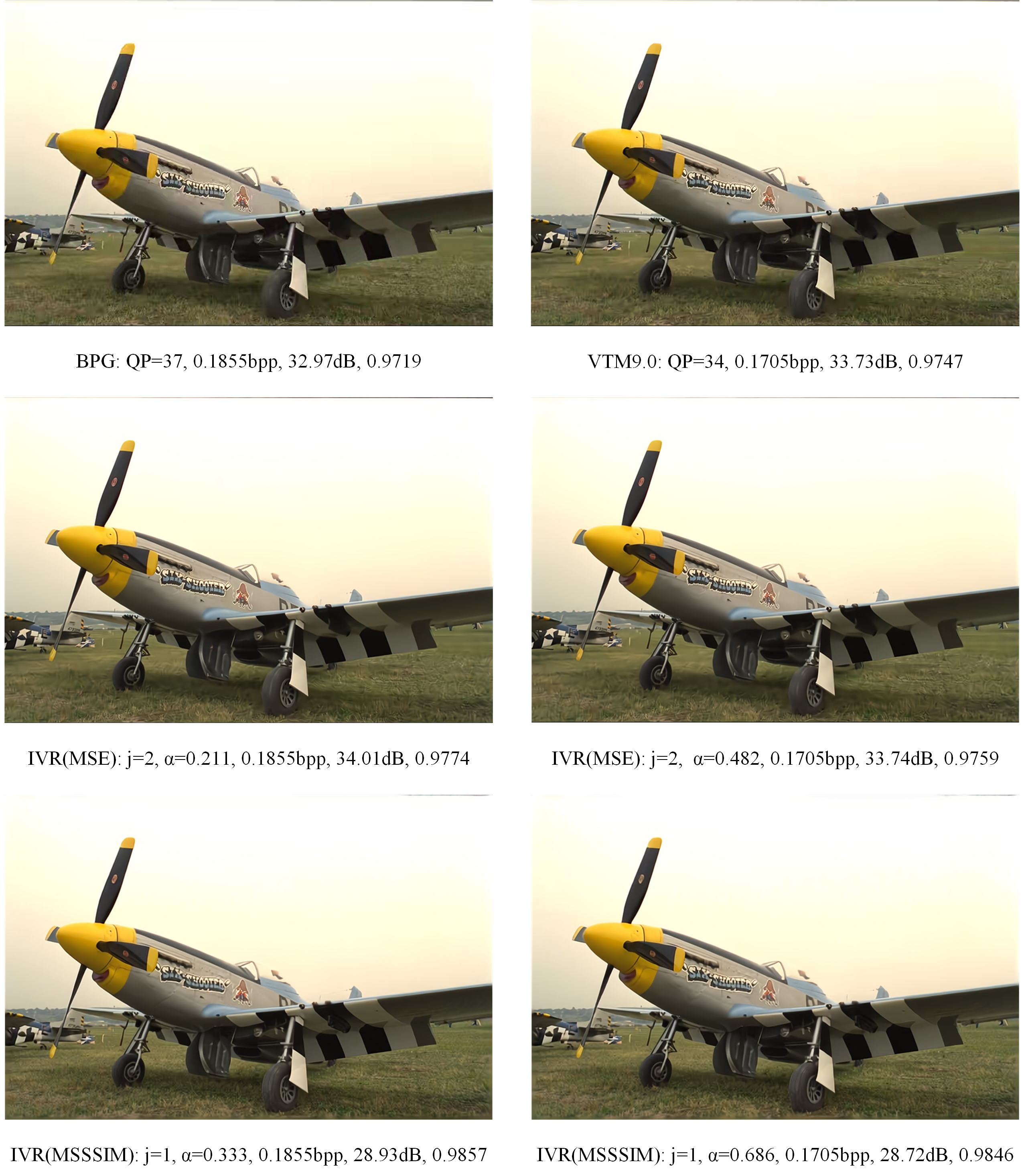}
 	\caption{Visualization of sample images (Kodim20 from Kodak dataset) reconstructed by our IVR networks, BPG and VTM 9.0. By adjusting $j$ and $\alpha$, the IVR networks (with Unet post-network) match the rates of BPG and VTM 9.0 with the fineness of 0.0001 BPP.}
 	\label{fig:visual2}
 \end{figure*}

 Figure~\ref{fig:visual1} and Figure~\ref{fig:visual2} show the decoded images of Kodim23 and Kodim20 by our IVR networks, VTM 9.0, and BPG. These figures reveal that the IVR method provides better reconstruction quality with fewer artifacts in terms of PSNR, MS-SSIM.

\end{document}